\documentclass[11pt]{article}

\usepackage{amsmath}
\usepackage{graphicx}
\usepackage{indentfirst}
\usepackage{amssymb}
\usepackage{cite}
\usepackage{color}
\usepackage{subfigure}
\usepackage{varwidth}
\usepackage{subfigure}
\usepackage[colorlinks=true, linkcolor=red, citecolor=blue, urlcolor=magenta]{hyperref}
\usepackage{xcolor}%参考文献
\usepackage{booktabs} % 表格

\begin{document}
\title{Polarization Images of Mini Boson Stars in Palatini $f(R)$ Gravity}

\date{}
\maketitle

\begin{center}
	\author{Xiao-Xiong Zeng,}$^{a}$
	\author{Chen-Yu Yang,}$^{b}$
	\author{Ke-Jian He,}$^{b}$
	\author{Yu-Xiang Huang,}$^{c}$
	\author{Guo-Ping Li,}$^{c}$
	\author{Sen Guo}$^{a}$\footnote{E-mail: sguophys@126.com (Corresponding author)}
	\\
	
	\vskip 0.25in
	$^{a}$\it{College of Physics and Electronic Engineering, Chongqing Normal University, Chongqing 401331, People's Republic of China}\\
	$^{b}$\it{Department of Mechanics, Chongqing Jiaotong University, Chongqing 400074, People's Republic of China}\\
	$^{c}$\it{Physics and Space College, China West Normal University, Nanchong 637000, People's Republic of China}\\
\end{center}

\vskip 0.6in
{\abstract
{ 
We investigate the optical properties of mini boson stars within the framework of Palatini $f(R)$ gravity, adopting the quadratic form $f(R) = R + \xi R^2$, where $\xi$ denotes the gravitational coupling constant. By deriving the modified scalar Lagrangian and numerically solving the corresponding field equations, we analyze photon trajectories and the resulting optical images produced by spherical light sources and thin accretion disks. Unlike Schwarzschild black holes, boson stars do not exhibit photon rings, as the derivative of the effective potential does not vanish anywhere. Consequently, their images are primarily shaped by direct emissions from photons completing a single orbit. We further investigate how the optical features and polarization images depend on the initial scalar field amplitude $\psi_0$ and the coupling parameter $\xi$. The numerical results, including effective potentials and detailed images, provide potential observational signatures that can be used to distinguish boson stars from black holes through high-resolution astronomical measurements.
}}

\thispagestyle{empty}
\newpage
\setcounter{page}{1}

%\tableofcontents%目录
%\newpage

%%%%%%%%%%%%%%%%%%%%%%%%%%%%%%%%%%%%%%%%%%%%%%%%%%%%%%%%%%%%%%%%%%%%%%%%
%%%%%%%%%%%%%%%%%%%%%%%%%%%%%%%%%%%%%%%%%%%%%%%%%%%%%%%%%%%%%%%%%%%%%%%%
%%%%%%%%%%%%%%%%%%%%%%%%%%%%%%%%%%%%%%%%%%%%%%%%%%%%%%%%%%%%%%%%%%%%%%%%
\section{Introduction}
In recent years, gravitational wave detections by LIGO and Virgo have provided unprecedented insights into the universe, facilitating in-depth studies of massive compact objects and advancing our understanding of binary black hole (BBH) formation mechanisms \cite{ligo2019gwtc,abbott2021gwtc}. These observations have placed constraints on the mass distribution of BBH systems, bridging the gap between the heaviest neutron stars and stellar-mass black holes (e.g., the GW190814 event \cite{abbott2020gw190814}), and have revealed significant discrepancies when compared to predictions from current stellar evolution models \cite{abbott2020gw190521}. Addressing these discrepancies between theoretical models and observational data has become a key challenge in contemporary astrophysics.

Compact objects with masses in the range of 50–100 $M_{\odot}$ present significant challenges to our current understanding of intermediate black hole masses \cite{abbott2020properties,de2021gw190521,bustillo2021gw190521}. Although Kerr black holes have traditionally been considered the primary progenitors of such events, the formation mechanisms leading to black holes in this mass range remain unclear. Recent studies suggest that the mergers of other massive, non-luminous compact objects could also account for these phenomena \cite{de2021gw190521,bustillo2021gw190521}. These alternatives, often referred to as "exotic compact objects" (ECOs) \cite{cardoso2019testing}, have garnered increasing attention, particularly in light of supporting evidence from electromagnetic observations \cite{herdeiro2021imitation}.

Among ECOs, boson stars hold a prominent position due to their unique theoretical characteristics. Composed of self-gravitating bosons, these ultracompact configurations exhibit masses and sizes ranging from atomic to astrophysical scales, depending on the boson mass. Since the pioneering works of Kaup \cite{kaup1968klein} and Ruffini and Bonazzola \cite{ruffini1969systems}, boson stars have been extensively studied, particularly in terms of their stability and dynamical behavior \cite{liebling2023dynamical,di2020dynamical}.

Numerical simulations have demonstrated that both rotating and non-rotating boson stars can attain masses on the order of several solar masses ($M_{\odot}$), consistent with astrophysical observations, while lacking an event horizon. Assuming spherical symmetry, boson stars can form binary systems whose gravitational wave emissions are comparable to those from BBH mergers \cite{palenzuela2007head,palenzuela2008orbital}. Beyond spherically symmetric models, equilibrium sequences of axisymmetric rotating boson stars have been successfully constructed \cite{schunck1998rotating,yoshida1997rotating,kleihaus2008rotating}, and the merger dynamics of binary orbiting boson stars have been extensively studied through simulations \cite{bezares2017final,palenzuela2017gravitational}.

Boson stars have long been proposed as viable alternatives to black holes, with some studies suggesting that the Galactic Center object, Sgr A*, may in fact be a boson star \cite{vincent2016imaging}. Under the thin accretion disk approximation, Rosa et al. studied the optical appearance of boson stars without incorporating gravitational coupling effects \cite{rosa2022shadows,rosa2023imaging}. Additionally, boson stars have been considered as central objects in accretion systems, replacing black holes in certain models \cite{guzman2006accretion}. Although the optical images of boson stars and black holes exhibit similarities, current observational techniques (under specific assumptions) are capable of distinguishing between them \cite{olivares2020tell}.

The shadow images of black holes under various accretion disk models have been extensively studied \cite{zeng2025holographic,guo2024image,cui2024optical,guo2024influence,he2024observational,hou2024unique,huang2024images,zhang2024imaging,zeng2020shadows,yang2024shadow,zeng2022shadows,yang2025shadow,wang2025imaging1,wang2025imaging2}. In contrast, this work focuses on the optical images of boson stars. To further explore their physical characteristics and imaging features, we adopt the Palatini $f(R)$ gravity theory \cite{olmo2011palatini,harko2018extensions}. This theory modifies general relativity by introducing a function $f(R)$ that depends on the scalar curvature $R$, thereby altering the dynamics of the gravitational field. Unlike traditional metric $f(R)$ theories, the Palatini approach treats the metric and affine connection as independent variables \cite{maso2021boson}, allowing for the incorporation of nonlinear correction terms within the relativistic framework. This approach offers a promising path for alternative gravitational theories, providing explanations for cosmological phenomena on large spatial scales without invoking dark matter or dark energy. It also establishes a theoretical foundation for resolving singularity issues and implementing modifications at high energy scales.

Incorporating $f(R)$ gravity into the study of boson stars requires deriving the corresponding field equations. Traditional formulations are based on Riemannian geometry, where spacetime is fully characterized by the metric tensor. In contrast, metric-affine geometry treats the metric and affine connection as independent variables \cite{olmo2011palatini,hehl1995metric,beltran2019geometrical}. Given the current uncertainty regarding the geometric structure of spacetime, both approaches warrant consideration. Under general relativity with minimal coupling to scalar fields, the choice of geometric framework typically leads to equivalent results. However, in $f(R)$ or more generalized theories, this choice results in distinct and inequivalent field equations. The nonlinear nature of the $f(R)$ Lagrangian ensures that the vacuum field equations reduce to the Einstein field equations with an effective cosmological constant, independent of the specific functional form of $f(R)$. From a computational perspective, the Palatini formalism offers a practical advantage: problems in modified gravity with minimal scalar coupling can be reformulated in terms of a modified scalar Lagrangian with minimal coupling, aligning with the structure of general relativity \cite{afonso2019correspondence,delhom2019ricci,afonso2018mapping,afonso2018mapping2}. This property has facilitated the derivation of analytical solutions for static, spherically symmetric scalar compact objects in Palatini $f(R)$ gravity and related frameworks \cite{afonso2019new}, providing a theoretical foundation for examining the impact of $f(R)$ gravity on the physical and observational characteristics of boson stars.

This paper adopts the quadratic form $f(R) = R + \xi R^2$, where $\xi$ denotes the gravitational coupling constant. The corresponding modified scalar Lagrangian is derived, and numerical methods are employed to compute photon trajectories and generate optical images for both spherical light sources and thin accretion disks. The structure of this work is organized as follows: Sec.~\ref{sub2} introduces the Palatini $f(R)$ framework and derives the associated field equations. Sec.~\ref{sub3} outlines the numerical methods. Sec.~\ref{sub4} applies ray-tracing techniques to analyze spherical light sources. Sec.~\ref{sub5} and Sec.~\ref{sub6} describe imaging under the thin accretion disk approximation and polarization images, respectively. Sec.~\ref{sub7} presents the numerical results, focusing on the influence of the scalar field $\psi_0$ and the coupling constant $\xi$ on the resulting images. Sec.~\ref{sub8} provides a summary of the entire study.

\section{Correspondence with general relativity and field equations}\label{sub2}

We start by examining the minimal coupling between the scalar field $\Psi$ and the metric $g_{ab}$, which is expressed as
\begin{equation}
	S_{f(R)} = \int d^4x \sqrt{-g} \frac{f(R)}{2\kappa} - \frac{1}{2} \int d^4x \sqrt{-g} P(X,\Psi), \label{frf}
\end{equation}
where $g$ is the determinant of the metric tensor $g_{ab}$, $\kappa = 8\pi$ is the gravitational coupling constant, $\Psi$ represents the complex scalar field, and $R = g^{ab} R_{ab}$ is the scalar curvature that governs gravitational dynamics, which can be modified by the Palatini $f(R)$ function. The term $P(X,\Psi)$ is given by $P(X,\Psi) = X - 2V(\Psi)$, with $X = g^{ab} \partial_a \bar{\Psi} \partial_b \Psi$ being the Lagrangian density for matter. The potential function $V(\Psi)$ is defined as
\begin{equation}
	V(\Psi) = -\frac{u^2}{2} \bar{\Psi} \Psi, \label{V}
\end{equation}
where $u$ is the mass of the scalar field. The boson star that arises from this potential term (\ref{V}) is referred to as a mini boson star. Throughout the paper, we adopt the geometric unit system, where $G = c = 1$.

In contrast to general relativity, in Palatini $f(R)$ gravity, the Christoffel symbols are not directly calculated from the metric but are determined through the following formula
\begin{equation}
	\Gamma^\sigma{}_{\mu\nu} = \frac{1}{2} \tilde{g}^{\sigma\rho} \left( \tilde{g}_{\rho\mu,\nu} + \tilde{g}_{\nu\rho,\mu} - \tilde{g}_{\mu\nu,\rho} \right),
\end{equation}
where
\begin{equation}
	\tilde{g}_{ab} \equiv f_R^{\prime} g_{ab}, \label{con}
\end{equation}
and $f_R^{\prime} = \frac{\partial f(R)}{\partial R}$ is known as the conformal factor. It is important to note that the conformal factor $f_R^{\prime}$ must be a function of the metric $g_{ab}$ and the energy-momentum tensor $T_{ab}$, satisfying the equation
\begin{equation}
	R f_R^{\prime} - 2 f(R) = \kappa T, \label{cfe}
\end{equation}
where $T$ is the trace of the energy-momentum tensor. The components of the energy-momentum tensor $T_{\mu\nu}$ are defined as
\begin{equation}
	T_{\mu\nu} \equiv -\frac{2}{\sqrt{-g}} \frac{\delta(\sqrt{-g} P(X, \Psi))}{\delta g^{\mu\nu}}.
\end{equation}
In general relativity, $f(R)$ corresponds to $R$. In this study, we assume that $f(R)$ takes the quadratic form
\begin{equation}
	f(R) = R + \xi R^2, \label{fr2}
\end{equation}
where $\xi$ is the gravitational coupling parameter. Substituting (\ref{fr2}) into (\ref{cfe}), we obtain $R = -\kappa T$, which is the same as the result obtained by taking the trace of the Einstein field equations. We refer to (\ref{fr2}) as the $f(R)$ framework.

Although Palatini $f(R)$ gravity is a modified theory of gravity, it has a correspondence with general relativity. Consider the following Einstein frame
\begin{equation}
	S_{E} = \int d^4x \sqrt{-\tilde{g}} \frac{\tilde{R}}{2\kappa} - \frac{1}{2} \int d^4x \sqrt{-\tilde{g}} K(Z, \Psi),
\end{equation}
where $\tilde{g}$ is the determinant of the metric $\tilde{g}_{ab}$, $K(Z, \Psi)$ is the Lagrangian minimally coupled to the metric $\tilde{g}_{ab}$, $Z = \tilde{g}^{ab} \partial_a \bar{\Psi} \partial_b \Psi$, and $\tilde{R} = \tilde{g}^{ab} \tilde{R}_{ab}$ is the scalar curvature calculated from $\tilde{g}_{ab}$.

It has been shown \cite{afonso2019correspondence} that for specific choices of $f(R)$ and $P(X, \Psi)$, there exists the relation
\begin{equation}
	K(Z, \Psi) = \frac{Z - \xi \kappa Z^2}{1 - 8 \xi \kappa V} - \frac{2V}{1 - 8 \xi \kappa V}, \label{KZc}
\end{equation}
which indicates that the nonlinear term $\xi$ in the $f(R)$ gravity framework is transferred to the Lagrangian in the Einstein frame. Due to this correspondence, we can solve the corresponding problem in general relativity using (\ref{KZc}) and then solve the Palatini $f(R)$ gravity problem. Once $\tilde{g}_{ab}$ and the scalar field $\Psi$ are obtained, the metric $g_{ab}$ can be determined by $\tilde{g}_{ab} \equiv f_R^{\prime} g_{ab}$.

For the Einstein frame, the components of the energy-momentum tensor are given by
\begin{align}
	\tilde{T}_{\mu\nu} &\equiv -\frac{2}{\sqrt{-\tilde{g}}} \frac{\delta(\sqrt{-\tilde{g}} K(Z, \Psi))}{\delta \tilde{g}^{\mu\nu}} \notag \\ 
	&= \frac{1}{2(1 + 4\xi \kappa u^2 |\Psi|^2)} \left[ (\partial_\mu \bar{\Psi} \partial_\nu \Psi + \partial_\nu \bar{\Psi} \partial_\mu \Psi) (1 - 2\xi \kappa Z) \right] \notag \\
	&\quad - \tilde{g}_{\mu\nu} \left( \partial^{\sigma} \bar{\Psi} \partial_{\sigma} \Psi (1 - \xi \kappa Z) + u^2 |\Psi|^2 \right). \label{eemt}
\end{align}
Next, we consider a spherically symmetric star described by the scalar field $\Psi(\tilde{r}, t) = \psi(\tilde{r}) e^{i\omega t}$ \cite{liebling2023dynamical, herdeiro2017asymptotically}, where $\omega$ is the oscillation frequency of the field. Both metrics $g_{ab}$ and $\tilde{g}_{ab}$ are assumed to be static and spherically symmetric
\begin{equation}
	ds_{f(R)}^2 = -A^2(r)dt^2 + B^2(r)dr^2 + r^2 d\theta^2 + r^2 \sin^2\theta d\phi^2,
\end{equation}
\begin{equation}
	ds_E^2 = -\tilde{A}^2(\tilde{r}) dt^2 + \tilde{B}^2(\tilde{r}) d\tilde{r}^2 + \tilde{r}^2 d\theta^2 + \tilde{r}^2 \sin^2\theta d\phi^2. \label{ele}
\end{equation}
From equations (\ref{eemt}) and (\ref{ele}), the components of the Einstein tensor $G_{tt}$ and $G_{\tilde{r} \tilde{r}}$ can be expressed in terms of the following logarithmic derivatives
\begin{align}
	\frac{\partial_{\tilde{r}} \tilde{B}}{\tilde{B}} &= \frac{1 - \tilde{B}^2}{2\tilde{r}} 
	+ \frac{1}{1 + 4\xi \kappa u^2 \psi^2} \frac{\kappa \tilde{r}}{4} \Bigg\{ u^2 \tilde{B}^2 \psi^2 \notag \\
	&\quad + \left( \omega^2 \psi^2 \frac{\tilde{B}^2}{\tilde{A}^2} + \varphi^2 \right) 
	\left( 1 - 2\kappa \xi \left( -\frac{\omega^2 \psi^2}{\tilde{A}^2} + \frac{\varphi^2}{\tilde{B}^2} \right) \right) \notag \\
	&\quad + 2\kappa \xi \tilde{B}^2 \left( \frac{\omega^2 \psi^2}{\tilde{A}^2} - \frac{\varphi^2}{\tilde{B}^2} \right)^2 \Bigg\}, \label{Grr}
\end{align}
\begin{align}
	\frac{\partial_{\tilde{r}} \tilde{A}}{\tilde{A}} &= \frac{\tilde{B}^2 - 1}{\tilde{r}} + \frac{\partial_{\tilde{r}} \tilde{B}}{\tilde{B}} 
	+ \frac{1}{1 + 4\xi \kappa u^2 \psi^2} \frac{\kappa \tilde{r}}{4} \left\{ -2u^2 \tilde{B}^2 \psi^2 \right. \notag \\
	&\quad - 2\kappa \xi \tilde{B}^2 \left( \frac{\omega^2 \psi^2}{\tilde{A}^2} - \frac{\varphi^2}{\tilde{B}^2} \right)^2 \Bigg\}, \label{Gtt}
\end{align}
where
\begin{equation}
	\varphi \equiv \partial_{\tilde{r}} \Psi(\tilde{r}, t),
\end{equation}
satisfying
\begin{align}
	\partial_{\tilde{r}} \varphi &= \frac{1}{(1 + 4\xi \kappa u^2 \psi^2)\left[1 - 2\xi \kappa \left(-\frac{\omega^2 \psi^2}{\tilde{A}^2} + \frac{3 \varphi^2}{\tilde{B}^2}\right)\right]} \Bigg\{ \notag \\
	&\quad - \varphi \left( \frac{2}{\tilde{r}} + \frac{\partial_{\tilde{r}} \tilde{A}}{\tilde{A}} - \frac{\partial_{\tilde{r}} \tilde{B}}{\tilde{B}} \right) \left(1 + 4\kappa \xi u^2 \psi^2 \right) \notag \\
	&\quad \times \left[1 - 2\xi \kappa \left(-\frac{\omega^2 \psi^2}{\tilde{A}^2} + \frac{\varphi^2}{\tilde{B}^2}\right)\right] \notag \\
	&\quad - \omega^2 \psi \frac{\tilde{B}^2}{\tilde{A}^2} \left[ 1 + 2\xi \kappa \left( \frac{\omega^2 \psi^2}{\tilde{A}^2} + \frac{\varphi^2}{\tilde{B}^2} \right) \right] \notag \\
	&\quad + \tilde{B}^2 \psi u^2 \left( 1 + 4\kappa \xi \frac{\varphi^2}{\tilde{B}^2} \right) \notag \\
	&\quad + \kappa \xi \left[\frac{4 \omega^2 \psi^2 \varphi}{\tilde{A}^2} \frac{\partial_{\tilde{r}} \tilde{A}}{\tilde{A}} \left( 1 + 4\kappa \xi u^2 \psi^2 \right) \right] \notag \\
	&\quad - \frac{4 \varphi^3}{\tilde{B}^2} \frac{\partial_{\tilde{r}} \tilde{B}}{\tilde{B}} \left(1 + 4\kappa \xi u^2 \psi^2 \right) \notag \\
	&\quad - 4 \kappa^2 \xi^2 u^2 \psi \tilde{B}^2 \left( \frac{\omega^4 \psi^4}{\tilde{A}^4} + \frac{3 \varphi^4}{\tilde{B}^4} \right) \Bigg\}. \label{phi}
\end{align}
The system of differential equations (\ref{Grr})-(\ref{phi}) represents the spherically symmetric boson star model in the quadratic $f(R)$ theory. The conformal factor $f_R^{\prime}$ takes the following form
\begin{equation}
	f_R^{\prime} = 1 + 2\xi R = 1 + 2\xi \kappa (X - 4V),
\end{equation}
noting that $X = f_R^{\prime} Z$, we can also express the conformal factor in terms of the variables of the Einstein frame as
\begin{equation}
	f_R^{\prime} = 1 + 2\xi \kappa \left[ \frac{(1 - 8\xi \kappa V) Z}{1 - 2\xi \kappa Z} - 4V \right].
\end{equation}

\section{Numerical analysis}\label{sub3}

To solve the system of differential equations (\ref{Grr})-(\ref{phi}), appropriate boundary conditions must be selected. We first establish the boundary conditions for the $f(R)$ framework. For physical consistency, we assume that the equations satisfy asymptotic flatness, similar to Schwarzschild spacetime at infinity, and regularity at the origin
\begin{align}
	\psi(\infty)=0,\quad\varphi(\infty)=0,\quad B^2(\infty)=1,\quad A^2(\infty)=1,\notag \\ \psi(0)=\psi_0,\quad\varphi(0)=0,\quad \partial_{r} B^2(0)=0,\quad\partial_{r} A^2(0)=0.
\end{align}
The asymptotic flatness condition also requires that $f_R^{\prime} \to 1$ as $r \to \infty$. According to equation~(\ref{con}), these conditions can be reformulated in terms of the Einstein frame variables. Therefore, the boundary conditions in the Einstein frame are as follows
\begin{align}
	\psi(\infty)&\equiv\psi(\tilde{r}(r))\Big|_{r=\infty}=0,\\\varphi(\infty)&\equiv\varphi(\tilde{r}(r))\Big|_{r=\infty}=0,\\\tilde{B}^2(\infty)&\equiv\tilde{B}^2(\tilde{r}(r))\Big|_{r=\infty}=f_R^{\prime}(\infty)B^2(\infty)=1,\\\tilde{A}^2(\infty)&\equiv\tilde{A}^2(\tilde{r}(r))\Big|_{r=\infty}=f_R^{\prime}(\infty)A^2(\infty)=1,\\\psi(0)&\equiv\psi(\tilde{r}(r))\Big|_{{r=0}}=\psi_{0},\\\varphi(0)&\equiv\varphi(\tilde r(r))|_{r=0}=0,\\\left[\partial_{{\tilde{r}}}\tilde{B}^{2}\right](0)&\equiv\left[\partial_{{\tilde{r}}}\tilde{B}^{2}(\tilde{r}(r))\right]_{{r=0}}=2\sqrt{f_R^{\prime}}B\partial_{{r}}(\sqrt{f_R^{\prime}}B)\partial_{{\tilde{r}}}r\Big|_{{r=0}}=\frac{B^{2}}{\sqrt{f_R^{\prime}}}\partial_{{r}}f_R^{\prime}\Big|_{{r=0}}=0,\label{btilde}\\\left[\partial_{{\tilde{r}}}\tilde{A}^{2}\right](0)&\equiv\left[\partial_{{\tilde{r}}}\tilde{A}^{2}(\tilde{r}(r))\right]_{{r=0}}=2\sqrt{f_R^{\prime}}A\partial_{{r}}(\sqrt{f_R^{\prime}}A)\partial_{{\tilde{r}}}r\Big|_{{r=0}}=\frac{A^{2}}{\sqrt{f_R^{\prime}}}\partial_{{r}}f_R^{\prime}\Big|_{{r=0}}=0.\label{atilde}
\end{align}
Substituting (\ref{btilde}) and (\ref{atilde}) into (\ref{Grr}) and (\ref{Gtt}), we find that $\tilde{B}^2(0) = 1$ and $\tilde{A}^2(0) = \tilde{A}_0^2$. This shows that if asymptotic flatness and regularity at the origin are assumed in the $f(R)$ framework, the same conditions will hold in the Einstein frame.

To make the equations dimensionless, we apply scale transformations to certain parameters to absorb their units
\begin{equation}
	r \to ur, \quad t \to \omega t.
\end{equation}
The coupling constant $\kappa$ in the Einstein field equations can be absorbed by redefining the matter fields
\begin{equation}
	\psi \to \sqrt{\frac{2}{\kappa}} \psi, \quad \varphi \to \sqrt{\frac{2}{\kappa}} \varphi,
\end{equation}
which makes the rescaled matter fields dimensionless. By exploiting the symmetry of the equations of motion, we can impose the following rescaling for the metric function
\begin{equation}
	\tilde{A} \to \frac{\omega}{u} \tilde{A}.
\end{equation}
For numerical calculations, the scalar field mass is set to $u = 1$. Similar to ideal fluid stars, boson stars can be characterized by their mass $m$ and radius $R$, both of which are related to the mass function $M(\tilde{r})$, defined as
\begin{equation}
	M(\tilde{r}) = \frac{\tilde{r}}{2} \left[ 1 - \tilde{B}^{-2}(\tilde{r}) \right].
\end{equation}
The total ADM mass is given by $m = M(\tilde{r} \to \infty)$, and the radius is taken as $R = 0.98m$. Since the mass of a boson star is highly concentrated, different methods of defining the radius yield only small discrepancies.

After selecting the boundary conditions and applying the scale transformations to the system of equations, numerical solutions can be derived. To avoid singularities, the origin $\tilde{r} = 0$ is replaced with a sufficiently small value of $\tilde{r}$. For a given central value of the scalar field $\psi_0$, the frequency $\omega$ must be adjusted to satisfy the boundary conditions. Notably, due to the scale transformation, $\omega$ is absorbed into $\tilde{A}$. The specific solution is obtained using a shooting method, integrating from the origin to the outer boundary. During the computation, multiple values of $\omega^{(n)}$ that satisfy the boundary conditions are identified, with the radial node number of $\psi$ increasing as $n$ increases. In this study, we focus on the nodeless case where $n = 0$, commonly referred to as the ground state or fundamental family.

%Through the scale transformations and redefinitions of parameters outlined in the previous section, the gravitational coupling constant $\xi$ acquires the dimensionality $[\xi] = [M]^2$, while the product $\xi u^2$ becomes dimensionless. Consequently, $\xi$ can be interpreted as a quantity measured in units of $u^2 = 1/l_u^2$. In physical contexts, it is commonly assumed that $\xi u^2 \sim l_\xi^2 / l_u^2 \ll 1$. To examine both positive and negative values of $\xi$, sufficiently large values are chosen to distinguish differences across various cases. In this study, we consider $\xi u^2 = \xi = -0.1, -0.05, 0.01, 0.1$, with $\xi = 0$ corresponding to general relativity.
%
%In the Palatini approach, the absolute bound on $\xi$ can be determined by analyzing the weak-field limit \cite{olmo2005gravity}, yielding the constraint $|\xi| \ll 2 \times 10^{12} \operatorname{cm}^2$. An alternative bound is derived by examining the equivalence between electromagnetic and Newtonian gravitational forces \cite{avelino2012eddington, jimenez2018born}, leading to the restriction $|\xi| < 6 \times 10^9 \operatorname{cm}^2$. In comparison, for the metric formulation, the bounds on $\xi$ range from $|\xi| < 5 \times 10^{15} \operatorname{cm}^2$ to $|\xi| < 1.7 \times 10^{18} \operatorname{cm}^2$. The former constraint is derived using data from the Gravity Probe B experiment, while the latter is based on precession data from binary pulsars. Furthermore, results from the Eöt-Wash experiment yield a significantly stricter bound of $|\xi| < 10^{-6} \operatorname{cm}^2$ \cite{naf20101}.

\section{Optical images with spherical light sources}\label{sub4}
In the previous sections, variables with $\sim$ were used to represent the Einstein frame, while variables without $\sim$ represented the $f(R)$ framework. From this section onward, we will exclusively use the Einstein frame, and the $\sim$ will be omitted from the variables for simplicity. For convenience, the line element (\ref{ele}) is rewritten in the following form:
\begin{equation}
	ds^2 = -A(r) dt^2 + B(r)^{-1} dr^2 + r^2 d\theta^2 + r^2 \sin^2\theta d\phi^2.
\end{equation}

This section focuses on the optical images of boson stars illuminated by a spherical light source. The backward ray-tracing method is employed, which assumes that photons are emitted from the observer and constructs pixel maps by solving the null geodesic equations numerically. The four-velocity of a photon is given by $ \left(\frac{\partial}{\partial x^{\mu}}\right)^a \dot{x}^\mu$, where $\tau$ is an affine parameter along the null geodesic, and $\cdot$ represents the derivative with respect to $\tau$. The motion of the photon satisfies the Euler-Lagrange equation
\begin{equation}
	\frac{d}{d\tau} \left( \frac{\partial K}{\partial \dot{x}^\mu} \right) = \frac{\partial K}{\partial x^\mu},
\end{equation}
where $K$ is the Lagrangian. Let $x^0 \equiv t$, $x^1 \equiv r$, $x^2 \equiv \theta$, and $x^3 \equiv \phi$. Then, $K$ can be expressed as
\begin{align}
	0 = K &= -\frac{1}{2} g_{\mu\nu} \dot{x}^{\mu} \dot{x}^{\nu} \notag \\
	&= -\frac{1}{2} \left[ -A(r) \dot{t}^2 + \frac{1}{B(r)} \dot{r}^2 + r^2 \dot{\theta}^2 + r^2 \sin^2\theta \dot{\phi}^2 \right]. \label{lag}
\end{align}
Since the metric components do not depend on $t$ and $\phi$, the spacetime admits two Killing vector fields, $\left(\frac{\partial}{\partial t}\right)^a$ and $\left(\frac{\partial}{\partial \phi}\right)^a$. By choosing a coordinate system such that $\theta$ is fixed along the null geodesic at $\pi/2$, two conserved quantities can be derived
\begin{align}
	\mathcal{E} &= \frac{\partial K}{\partial \dot{t}} = A(r) \frac{dt}{d\tau}, \label{lage} \\
	\mathcal{L} &= -\frac{\partial K}{\partial \dot{\phi}} = r^2 \frac{d\phi}{d\tau}. \label{lagl}
\end{align}
Using equations~(\ref{lag})--(\ref{lagl}), the remaining components of the four-velocity can be obtained as
\begin{align}
	\dot{t} &= \frac{1}{I A(r)}, \label{geo1} \\
	\dot{r} &= \sqrt{\frac{1}{I^2} \frac{B(r)}{A(r)} - \frac{1}{r^2} B(r)}, \\
	\dot{\phi} &= \pm \frac{1}{r^2}, \label{geo3}
\end{align}
where the "$+$" sign indicates the clockwise direction and "$-$" indicates the counterclockwise direction. The impact parameter $I$ is defined as
\begin{equation}
	I \equiv \frac{|\mathcal{L}|}{\mathcal{E}}. \label{impara}
\end{equation}
In this study, we consider the effective potential as
\begin{equation}
	V_{\mathrm{eff}} = \dot{r}^2. \label{veff}
\end{equation}
If $V_{\mathrm{eff}} = V_{\mathrm{eff}}^{\prime} = 0$ (where "$^{\prime}$" denotes the derivative with respect to the radial coordinate $r$), a photon ring forms around the boson star. The stability of photon orbits can be determined by the second derivative of the effective potential
\begin{align}
	V_{\mathrm{eff}}^{\prime\prime} 
	\begin{cases} 
		< 0, & \text{unstable}, \\ 
		= 0, & \text{marginally stable}, \\ 
		> 0, & \text{stable},
	\end{cases} \label{veffd2}
\end{align}
where the marginally stable orbit is analogous to the innermost stable circular orbit (ISCO) in black hole spacetimes~\cite{li2025observational}.

The equations (\ref{geo1})-(\ref{geo3}) are first-order differential equations for photon geodesics. To determine the photon trajectory, it is necessary to fix the integration constants. For this purpose, we choose an observer. Considering the arbitrariness of the observer's spacetime location, we adopt a zero-angular-momentum observer (ZAMO). Assuming the observer is located at $(t_o, r_o, \theta_o, \phi_o)$, a locally orthonormal tetrad can be constructed in the observer's vicinity as follows
\begin{align}
	(\hat{e}_0)^a &= \frac{1}{\sqrt{-g_{tt}}} \left( \frac{\partial}{\partial t} \right)^a, \quad (\hat{e}_1)^a = \frac{1}{\sqrt{g_{rr}}} \left( \frac{\partial}{\partial r} \right)^a, \notag \\
	(\hat{e}_2)^a &= \frac{1}{\sqrt{g_{\theta\theta}}} \left( \frac{\partial}{\partial \theta} \right)^a, \quad (\hat{e}_3)^a = \frac{1}{\sqrt{g_{\phi\phi}}} \left( \frac{\partial}{\partial \phi} \right)^a. \label{frame}
\end{align}
It should be noted that the choice of tetrad is not unique, and different tetrads are related by Lorentz transformations.

To describe the direction of light as observed, we introduce celestial coordinates. Let $\overrightarrow{OA}$ denote the three-momentum of the photon, geometrically interpreted as the projection of the tangent vector of the photon geodesic at point $O$ onto the observer's three-dimensional subspace. The celestial sphere is defined as a sphere centered at $O$ with a radius of $|\overrightarrow{OA}|$. We define $\Theta$ as the angle between $\overrightarrow{OA}$ and $(\hat{e}_1)^a$, and $\Phi$ as the angle between $\overrightarrow{OA}$ and $(\hat{e}_2)^a$. The celestial coordinate system is then expressed in terms of $(\Theta, \Phi)$. Using celestial coordinates, the tangent vector of the photon geodesic in the tetrad (\ref{frame}) can be written as
\begin{align}
	U^a = |\overrightarrow{OA}| \left[ -(\hat{e}_0)^a + \cos\Theta (\hat{e}_1)^a + \sin\Theta \cos\Phi (\hat{e}_2)^a + \sin\Theta \sin\Phi (\hat{e}_3)^a \right], \label{geotan}
\end{align}
where the negative sign in front of $-(\hat{e}_0)^a$ ensures that the tangent vector is a past-directed null vector. Furthermore, since the photon trajectory is independent of its energy, we normalize the photon energy to 1, i.e., $|\overrightarrow{OA}| = 1$.

On the other hand, for each light ray, the coordinates $(t, r, \theta, \phi)$ are functions of the affine parameter $\tau$, and its tangent vector takes the general form
\begin{equation}
	U^a = \dot{t} \left( \frac{\partial}{\partial t} \right)^a + \dot{r} \left( \frac{\partial}{\partial r} \right)^a + \dot{\theta} \left( \frac{\partial}{\partial \theta} \right)^a + \dot{\phi} \left( \frac{\partial}{\partial \phi} \right)^a. \label{gentan}
\end{equation}
From equations (\ref{frame})-(\ref{gentan}), the photon four-momentum and celestial coordinates are found to have a one-to-one correspondence. Given the photon four-momentum, the celestial coordinates can be uniquely determined. Conversely, if the celestial coordinates are specified, the photon four-momentum can be obtained through coordinate transformations. Thus, in combination with the observer's position, the initial conditions for the photon's equations of motion, $\left.(x^\mu, p_\mu)\right|_o$, can be completely specified.

To obtain the optical image of the boson star, it is necessary to map the celestial coordinates $(\Theta, \Phi)$ to the Cartesian coordinates $(x, y)$ on the imaging plane. This mapping depends on the choice of the camera model. In this study, we adopt a wide-angle fisheye camera model. A Cartesian coordinate system is established on the imaging plane with $O^\prime$ as the origin. The projection coordinates of point $A$ on the plane are given by
\begin{equation}
	x_A = -2 \tan\left(\frac{\Theta}{2}\right) \sin\Phi, \quad y_A = -2 \tan\left(\frac{\Theta}{2}\right) \cos\Phi. \label{proco}
\end{equation}
Next, we discuss the camera projection. The field of view angle $\alpha_{\text{fov}}$ determines the range of the camera's vision. For simplicity, we take $\alpha_{\text{fov}}/2$ in both the $x$ and $y$ directions, thereby defining a square screen. The side length of the screen is given by
\begin{equation}
	L = 2 \left|\overrightarrow{OA}\right| \tan \frac{\alpha_{\text{fov}}}{2}.
\end{equation}
On the imaging plane, we divide the screen into an $n \times n$ grid of pixels, where each pixel has a side length of
\begin{equation}
	l = \frac{L}{2} = \frac{2}{n} \left|\overrightarrow{OA}\right| \tan \frac{\alpha_{\text{fov}}}{2}.
\end{equation}
The center of each pixel is labeled with coordinates $(i, j)$, where the bottom-left pixel is $(1, 1)$, and the top-right pixel is $(n, n)$. The indices $i$ and $j$ range from $1$ to $n$ (in this study, $n = 300$). The relationship between the pixel coordinates $(i, j)$ and the projection coordinates $(x_A, y_A)$ is given by
\begin{equation}
	x_A = l \left(i - \frac{n+1}{2}\right), \quad y_A = l \left(j - \frac{n+1}{2}\right). \label{proco2}
\end{equation}
By comparing equations (\ref{proco}) and (\ref{proco2}), the relationship between the pixel coordinates $(i, j)$ and the celestial coordinates $(\Theta, \Phi)$ can be expressed as
\begin{align}
	\tan\Phi &= \frac{2j - (n+1)}{2i - (n+1)}, \notag \\
	\tan\frac{\Theta}{2} &= \frac{1}{n} \tan\left(\frac{\alpha_{\text{fov}}}{2}\right) \sqrt{\left(i - \frac{n+1}{2}\right)^2 + \left(j - \frac{n+1}{2}\right)^2}.
\end{align}

\section{Optical images with thin accretion disk}\label{sub5}

This section investigates the optical images of boson stars illuminated by an accretion disk. Similar to black holes, boson stars may host accretion disks under realistic astrophysical conditions. However, in contrast to black holes, boson stars lack an event horizon, which prevents accreting matter from falling inward. The accretion disk is assumed to lie on the equatorial plane, remaining stationary, axisymmetric, and both optically and geometrically thin. It emits radiation that reaches the zero-angular-momentum observer (ZAMO). Typically, when light rays interact with the accretion disk, fluctuations in light intensity occur due to photon emission and absorption. For simplicity, we neglect refraction effects in this model. Under these assumptions, the variation in light intensity is governed by the following equation
\begin{equation}
	\frac{d}{d\tau} \left( \frac{S_\nu}{\nu^3} \right) = \frac{E_\nu - A_\nu S_\nu}{\nu^2}, \label{ligstr}
\end{equation}
where $\tau$ denotes the affine parameter along the null geodesic, and $S_\nu$, $E_\nu$, and $A_\nu$ represent the specific intensity, emissivity, and absorption coefficient at frequency $\nu$, respectively. In the absence of photon absorption and emission, the quantity $S_\nu / \nu^3$ is conserved along the geodesics.

In the thin disk approximation, the accretion disk lies in the equatorial plane, so outside the equatorial plane, $E_\nu = A_\nu = 0$. Under this condition, the total light intensity at each position on the observer's screen is given by
\begin{equation}
	Q_0 = \sum\limits_{n=1}^{N_{\max}} f_n (\chi_n)^3 E_n,
\end{equation}
where $n = 1, \dots, N_{\max}$ represents the number of times a ray crosses the equatorial plane, $f_n$ is a fudge factor determined by the specific accretion disk model, and $\chi_n \equiv \frac{\nu_0}{\nu_n}$ is the redshift factor.

From the above equation, it is evident that the intensity $Q_0$ depends on the factors $f_n$, $\chi_n$, and $E_n$. The fudge factor $f_n$ primarily influences the intensity of the narrow photon ring and has a minimal impact on the overall image. Therefore, for simplicity, we set $f_n = 1$. There are various models available for the emissivity $E$ of black holes. To match astronomical observations (e.g., images of M87$^{\star}$ and Sgr A$^{\star}$), $E$ is commonly assumed to follow a second-order polynomial in log-space. In this study, we adopt the Gralla-Lupsasca-Marrone model \cite{gralla2020shape}, expressed as
\begin{equation}
	E = \frac{\exp \left[ -\frac{1}{2} \left( \gamma + \operatorname{arcsinh} \left( \frac{r - \beta}{\sigma} \right) \right)^2 \right]}{\sqrt{(r - \beta)^2 + \sigma^2}}, \label{e}
\end{equation}
which remains widely accepted due to its consistency with predictions from general relativistic magnetohydrodynamic simulations of astrophysical accretion disks \cite{vincent2022images}. In equation (\ref{e}), the parameters $\gamma$, $\beta$, and $\sigma$ represent the rate of increase, radial translation, and dilation of the profile, respectively. These parameters determine the shape of the emitted photon distribution and can be adjusted to match numerical simulations with observations. For simplicity, we set $\beta = 6m$, $\sigma = m$, and $\gamma = 0$ in this study, where $m$ is the mass of the boson star.

Next, we determine the redshift factor $\chi_n$. In this model, the accretion flow is treated as an electrically neutral plasma traveling along timelike geodesics. As discussed earlier, these geodesics are associated with two conserved quantities, $E_{tl}$ and $L_{tl}$. Outside the marginally stable orbit, the flow follows circular orbits, and the angular velocity at radius $r_n$ is given by
\begin{equation}
	\Omega_n = \frac{u^\phi}{u^t} \bigg|_{r = r_n}.
\end{equation}
The redshift factor $\chi_n$ can be expressed as
\begin{equation}
	\chi_n = -\frac{z}{Z(1 - I \Omega_n)},
\end{equation}
where $I$ is the photon's impact parameter, defined in equation (\ref{impara}), and $z$ is the ratio of the energy observed on the screen to the conserved quantity $E$ given in equation (\ref{lage}). In an asymptotically flat spacetime, when the observer is at infinity, $z = 1$. The quantity $Z$ is defined as
\begin{equation}
	Z \equiv \sqrt{\frac{1}{g_{tt} + g_{\phi\phi} \Omega_n^2}} \Bigg|_{r = r_n}.
\end{equation}
Through this analysis, the intensity and redshift factor for each pixel on the screen can be quantitatively computed.

\section{Polarized Images}\label{sub6}
The previous section introduced the shadow images of boson stars under the thin accretion disk model. Based on this, this section will study the polarization images of boson stars, aiming to gain a more comprehensive understanding of their radiation characteristics. In this paper, it is assumed that the emitted polarization light originates from synchrotron radiation produced by electrons in the plasma. For observers comoving with the plasma, the polarization direction $\vec{f}$ of the emitted light is always perpendicular to both the local magnetic field $\vec{b}$ and the photon wavevector $\vec{k}$
\begin{equation}
	\vec{f} = \frac{\vec{k} \times \vec{b}}{|\vec{k}| |\vec{b}|}.
\end{equation}
The general covariant form of the above equation can be written as
\begin{equation}
	f^\mu \propto \epsilon^{\mu\nu\alpha\beta} u_\nu k_\alpha b_\beta,
\end{equation}
where $u^\mu$ is the four-velocity of the observer and $\epsilon^{\mu\nu\alpha\beta}$ is the Levi-Civita tensor. The intensity of the emitted linear polarized light and natural light is represented by the emission functions $E_p$ and $E_i$, respectively. To simplify the calculation, we assume that the emission intensity only depends on the radial coordinate and is independent of photon frequency and magnetic field, so we have
\begin{equation}
	E_i = E_i(r), \qquad E_p = \mathcal{C} E_i(r),
\end{equation}
where $\mathcal{C} \in [0, 1]$ determines the proportion of the emitted light that is linearly polarized. If $\mathcal{C} = 1$, it indicates that the emitted light is fully linearly polarized. Under the geometric optics approximation, the polarization vector $f^\mu$ undergoes parallel transport along the photon geodesic
\begin{equation}
	k^\nu \nabla_\nu f^\mu = 0,
\end{equation}
or equivalently,
\begin{equation}
	\frac{df^\mu}{d\tau} + \Gamma^\mu_{\nu\alpha} k^\nu f^\alpha = 0,
\end{equation}
where $\tau$ is the affine parameter of the photon. At the observer's location, the linear polarization intensity $P_{\nu_o}$ and total intensity $Q_{\nu_o}$ are the same as in the unpolarized case
\begin{equation}
	P_{\nu_o} = \chi^3 E_p, \qquad Q_{\nu_o} = \chi^3 E_i,
\end{equation}
where $\chi$ is the redshift factor. As discussed previously, we have established a set of orthogonal tetrads (ZAMO) at the observer's location and set up the imaging plane. Based on this, a set of orthonormal vectors $(e_{(\theta)},\,e_{(\phi)})$ is chosen as the basis, and the projection of $f^\mu$ onto the imaging plane can be expressed as
\begin{equation}
	f^{(\alpha)} = f^\mu \cdot e_\alpha = -f^\mu \cdot e_\phi, \qquad f^{(\beta)} = f^\mu \cdot e_\beta = -f^\mu \cdot e_\theta.
\end{equation}
The Stokes parameters $\mathsf{Q}$ and $\mathsf{U}$ are commonly used to describe the polarization state. According to \cite{huang2024coport}, $\mathsf{Q}$ and $\mathsf{U}$ satisfy the principle of linear superposition, so
\begin{equation}
	\mathsf{Q}_{all} = \sum_{n=1}^{N} \chi_n^3 E_{pn} \left[ \left(f_n^{(\alpha)}\right)^2 - \left(f_n^{(\beta)}\right)^2 \right], \quad
	\mathsf{U}_{all} = \sum_{n=1}^{N} \chi_n^3 E_{pn} \left( 2 f_n^{(\alpha)} f_n^{(\beta)} \right),
\end{equation}
where $N_{\max}$ represents the number of times the photon passes through the equatorial plane. Using the Stokes parameters, the total linear polarization intensity and the electric vector position angle (EVPA) can be further calculated
\begin{equation}
	\mathsf{P}_o = \left( \mathsf{Q}_{all}^2 + \mathsf{U}_{all}^2 \right)^{1/2}, \quad \vartheta = \frac{1}{2} \arctan \left( \frac{\mathsf{U}_{all}}{\mathsf{Q}_{all}} \right). \label{eq:f}
\end{equation}
We adopt the gauge $f^{(\beta)} > 0, \vartheta \in (0, \pi)$. Equation (\ref{eq:f}) corresponds to the magnitude and direction of the polarization vector $\vec{f}$. Additionally, the polarization vector $\vec{f}$ must satisfy the normalization condition
\begin{equation}
	f^\mu f_\mu = 1.
\end{equation}
Based on this, together with the thin accretion disk model discussed earlier, we can calculate the complete polarization images.

\section{Numerical results}\label{sub7}
Using the theoretical framework outlined above, we apply numerical methods to solve for boson star configurations and examine their optical images under various light source conditions. The solutions depend on the initial scalar field $\psi_0$ and the gravitational coupling parameter $\xi$ within the context of Palatini gravity. We begin by fixing $\xi$ and varying $\psi_0$ to investigate how the initial scalar field influences the optical appearance of boson stars. Then, with $\psi_0$ held constant, we explore the impact of the coupling parameter $\xi$ on their optical properties. Specifically, we compute numerical solutions for the scalar field and the metric components as functions of the radial coordinate $r$. We then construct fitting functions for the metric components based on selected functional forms. These fitted metrics are subsequently used to analyze the optical images of boson stars illuminated by spherical light sources and thin accretion disks.

\subsection{Varying scalar field}\label{6.1}

\begin{figure}[htbp]
	\centering
	\includegraphics[width=.4\textwidth]{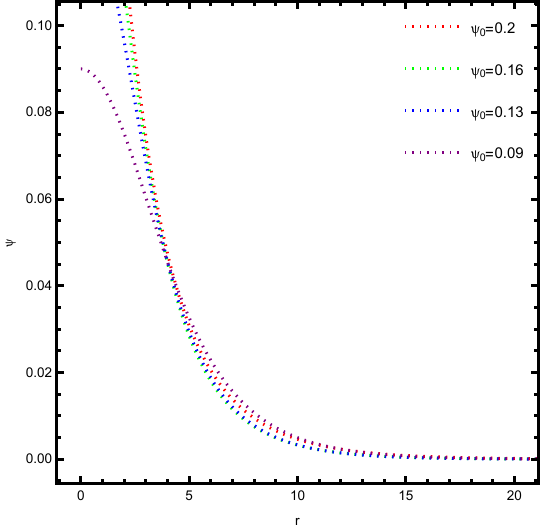}
	\caption{Radial variation of the scalar field for different boson stars, with $\xi=0.05$. \label{fig1}}
\end{figure}

\begin{figure}[htbp]
	\centering
	\includegraphics[width=.4\textwidth]{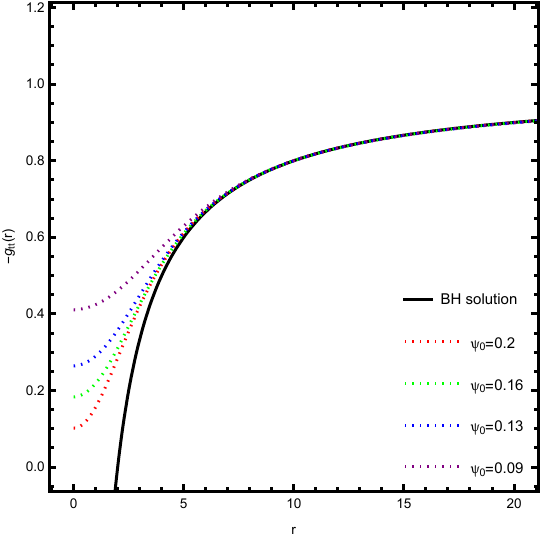}
	\qquad
	\includegraphics[width=.4\textwidth]{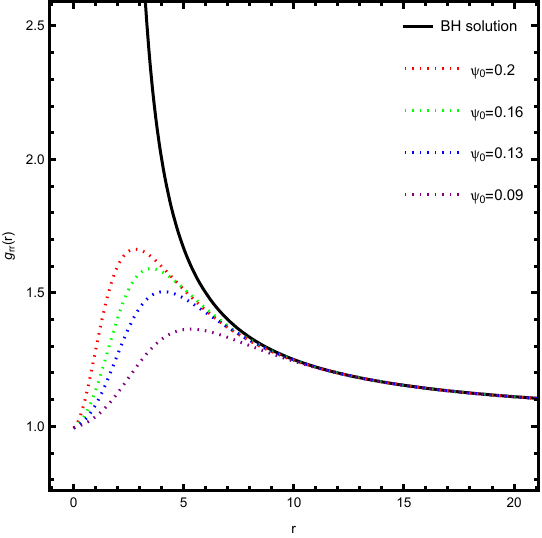}
	\caption{Metric components $-g_{tt}$ (left) and $g_{rr}$ (right) for different boson stars and the Schwarzschild black hole, with $\xi=0.05$. The mass of the Schwarzschild black hole is set to $1$. The black solid line represents the Schwarzschild black hole, while the dotted lines with different colors represent different boson stars. \label{fig2}}
\end{figure}

In this subsection, we fix $\xi = 0.05$ and consider the values $\psi_0 = 0.2, 0.16, 0.13, 0.09$. Figure~\ref{fig1} illustrates the variation of the scalar field $\psi$ as a function of the radial distance $r$ for different values of $\psi_0$. It is observed that $\psi$ decreases rapidly with increasing $r$ and tends to zero as $r \to \infty$. Figure~\ref{fig2} presents the numerical solutions for the metric components $-g_{tt}$ and $g_{rr}$. For comparison, the corresponding components of the Schwarzschild black hole metric are also shown. Unlike the Schwarzschild black hole, the boson star does not possess an event horizon, as evidenced by the fact that the metric components remain finite. In contrast, the Schwarzschild black hole exhibits a divergence at the event horizon. Nevertheless, the asymptotic behavior of the boson star matches that of the Schwarzschild solution. For various initial conditions $\psi_0$, both $-g_{tt}$ and $g_{rr}$ approach 1 as $r \to \infty$. When the Schwarzschild black hole mass is set to $M = 1$, all solutions converge for $r > 10M$, indicating that the spacetime is asymptotically flat.

After obtaining the numerical solutions for the metric components, we can theoretically study the optical image of the boson star using numerical methods. However, due to the presence of infinity in the numerical solutions, we cannot directly use the numerical metric for calculations. Therefore, we employ two specific functions to fit the numerical metric components. The fitting process involves minimizing the absolute value of the loss function and ensuring that the fitted metric asymptotically approaches the Schwarzschild metric as $r \to \infty$. We find that the following functions provide an almost perfect fit
\begin{align}
	g_{tt} &= -\exp\left[\alpha_7 \left( \exp \left( -\frac{1 + \alpha_1 r + \alpha_2 r^2}{\alpha_3 + \alpha_4 r + \alpha_5 r^2 + \alpha_6 r^3} \right) - 1 \right) \right], \label{gtt} \\
	g_{rr} &= \exp\left[\beta_7 \left( \exp \left( -\frac{1 + \beta_1 r + \beta_2 r^2}{\beta_3 + \beta_4 r + \beta_5 r^2 + \beta_6 r^3} \right) - 1 \right) \right]. \label{grr}
\end{align}

\begin{figure}[htbp]
	\centering
	\includegraphics[width=.4\textwidth]{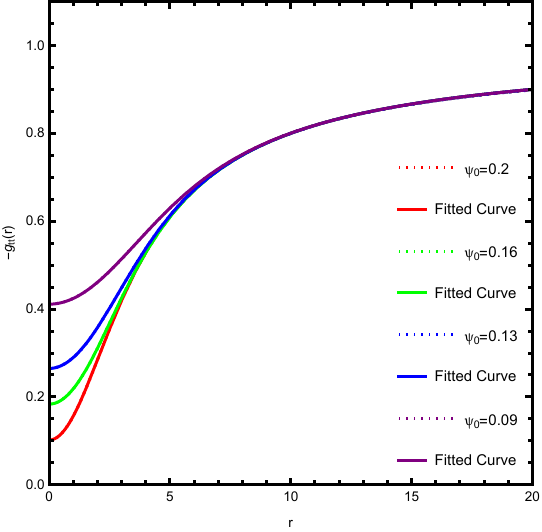}
	\qquad
	\includegraphics[width=.4\textwidth]{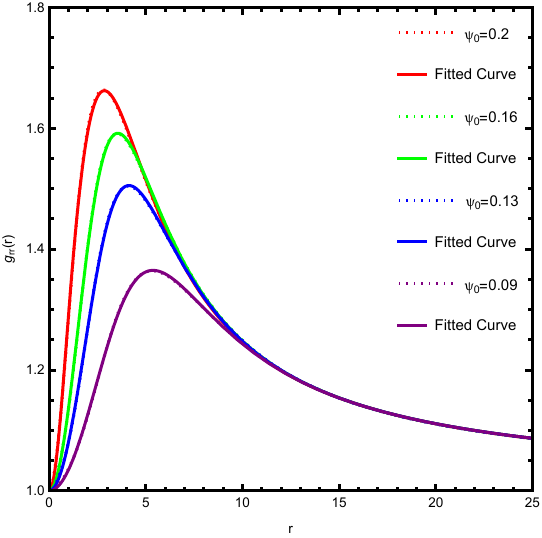}
	\caption{Numerical results and fitting functions of the metric components $-g_{tt}$ (left) and $g_{rr}$ (right) for different boson stars, with $\xi=0.05$. The dotted lines represent the numerical results, while the solid lines represent the fitting functions. \label{fig3}}
\end{figure}

\begin{figure}[htbp]
	\centering
	\includegraphics[width=.5\textwidth]{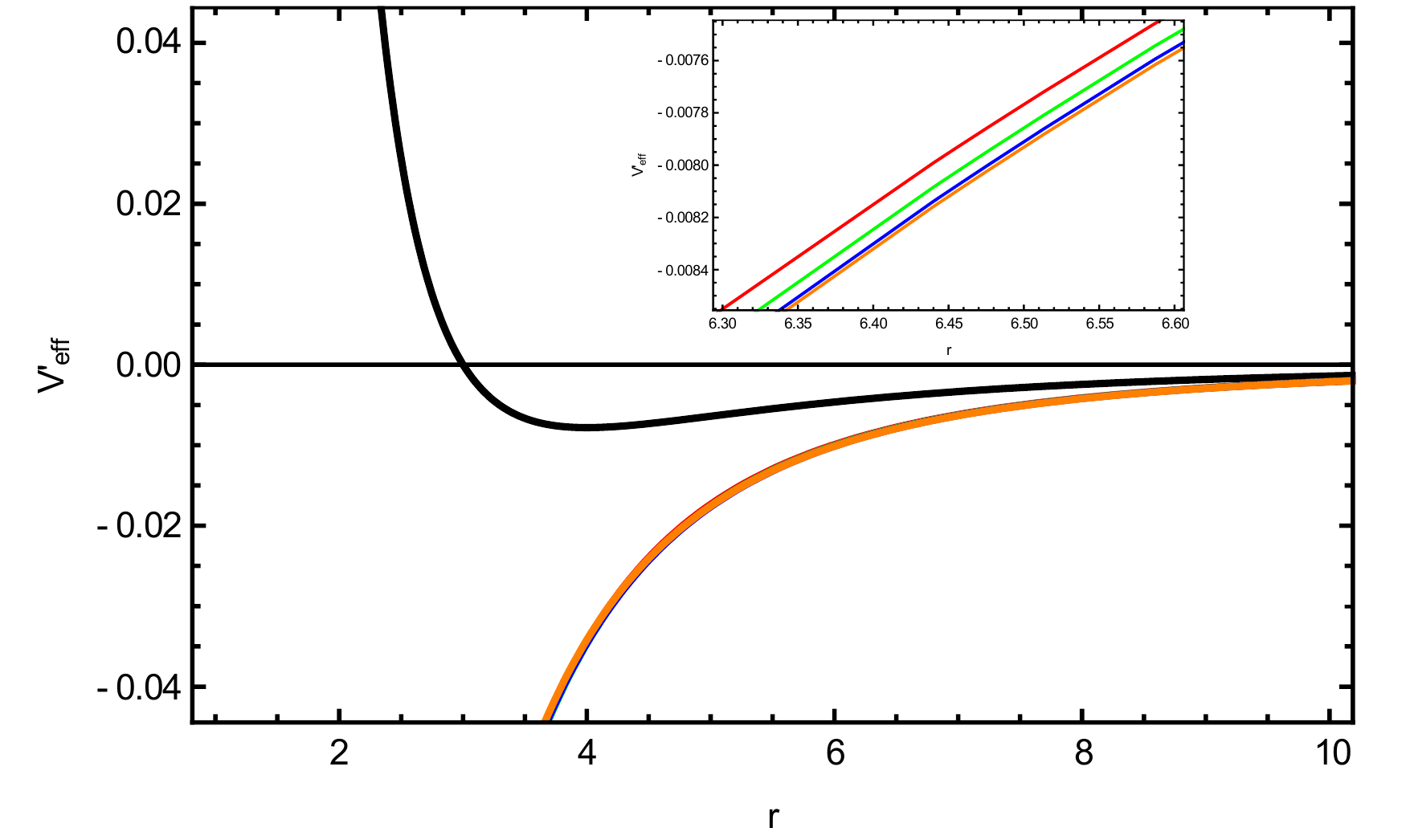}
	\caption{Derivative of the effective potential. The red, blue, green, and orange solid lines correspond to $\psi_0 = 0.2$, $0.16$, $0.13$, and $0.09$, with $\xi = 0.05$ fixed. The black solid line represents a Schwarzschild black hole with mass $1$.\label{fig4}}
\end{figure}

The fitting results are shown in Figure~\ref{fig3}, where the numerical results are represented by dotted lines and the fitting curves by solid lines. It can be observed that for both $-g_{tt}$ and $g_{rr}$, the numerical results almost perfectly coincide with the fitting results. Tables~\ref{tab1} and~\ref{tab2} present the estimated values of the fitting function parameters and the boson star mass $m$ for four initial conditions. It is evident that as the initial value $\psi_0$ increases, the mass $m$ decreases. Figure~\ref{fig4} shows the derivative of the effective potential with respect to the radial coordinate $r$. For the Schwarzschild black hole, $V_{\mathrm{eff}}^{\prime}(3) = 0$, indicating the presence of a photon ring. In contrast, for boson stars, $V_{\mathrm{eff}}^{\prime}(r)$ remains negative for all values of $\psi_0$, implying the absence of photon rings. This feature is also evident in Figure~\ref{fig6}. From the above discussion, we conclude that if the optical image of a compact object does not contain a photon ring, the object is unlikely to be a black hole and may instead be a boson star.

\begin{table*}[htbp]
	\centering
	\caption{\label{tab1}Estimated values of the parameters $\alpha_i$ $(i=1,2,\dots,7)$ 
		in the fitting function $g_{tt}$ [see equation~(\ref{gtt})], where $m$ denotes the mass 
		of the boson star and $\xi=0.05$ is fixed.}
	\begin{tabular}{lccccccccc}
		\toprule
		Type & $\psi_0$ & $m$ & $\alpha_1$ & $\alpha_2$ & $\alpha_3$ & $\alpha_4$ & $\alpha_5$ & $\alpha_6$ & $\alpha_7$ \\
		\midrule
		BS1 & 0.20 & 0.442 & 0.803 & 0.330 & -0.274 & -0.218 & 0.202  & -0.023 & -0.061 \\
		BS2 & 0.16 & 0.528 & 0.140 & 0.192 & -0.311 & -0.046 & -0.091 & -0.013 & -0.071 \\
		BS3 & 0.13 & 0.581 & 0.039 & 0.118 & -0.309 & -0.014 & -0.051 & -0.006 & -0.055 \\
		BS4 & 0.09 & 0.625 & 0.006 & 0.056 & -0.300 & -0.003 & -0.023 & -0.001 & -0.033 \\
		\bottomrule
	\end{tabular}
\end{table*}

\begin{table*}[htbp]
	\centering
	\caption{\label{tab2}Estimated values of the parameters $\beta_i$ $(i=1,2,\dots,7)$ 
		in the fitting function $g_{rr}$ [see equation~(\ref{grr})], where $m$ denotes the mass 
		of the boson star and $\xi=0.05$ is fixed.}
	\begin{tabular}{lccccccccc}
		\toprule
		Type & $\psi_0$ & $m$ & $\beta_1$ & $\beta_2$ & $\beta_3$ & $\beta_4$ & $\beta_5$ & $\beta_6$ & $\beta_7$ \\
		\midrule
		BS1 & 0.20 & 0.442 & -30.828 & -2.965 & 2.026 & 5.501 & 2.260 & 0.052 & -0.018 \\
		BS2 & 0.16 & 0.528 & -16.192 & -2.458 & 3.725 & 1.750 & 1.833 & 0.068 & -0.032 \\
		BS3 & 0.13 & 0.581 & -11.720 & -1.710 & 4.451 & 0.714 & 1.314 & 0.043 & -0.033 \\
		BS4 & 0.09 & 0.625 &  -8.136 & -0.746 & 4.543 & 0.319 & 0.660 & 0.011 & -0.026 \\
		\bottomrule
	\end{tabular}
\end{table*}

\begin{figure*}[htbp]
	\centering
	
	\subfigure[\scriptsize $\psi_0=0.2$]{
		\includegraphics[width=0.22\textwidth]{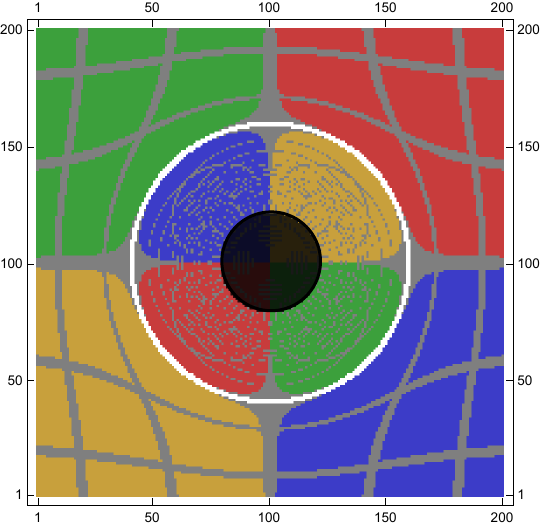}
	}
	\hfill
	\subfigure[\scriptsize $\psi_0=0.16$]{
		\includegraphics[width=0.22\textwidth]{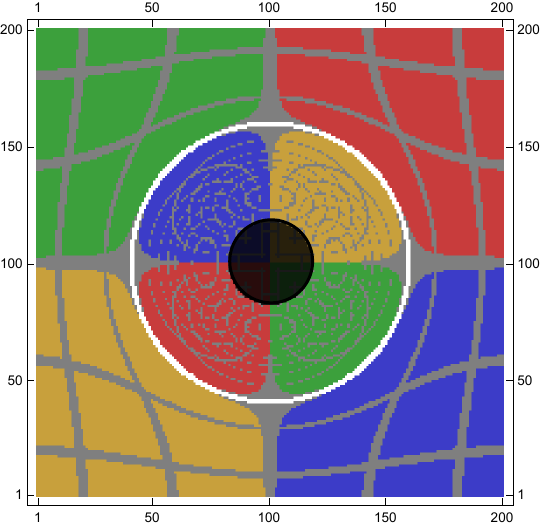}
	}
	\hfill
	\subfigure[\scriptsize $\psi_0=0.13$]{
		\includegraphics[width=0.22\textwidth]{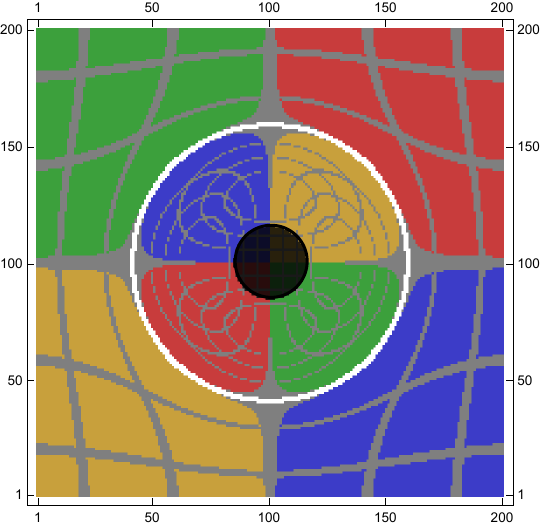}
	}
	\hfill
	\subfigure[\scriptsize $\psi_0=0.09$]{
		\includegraphics[width=0.22\textwidth]{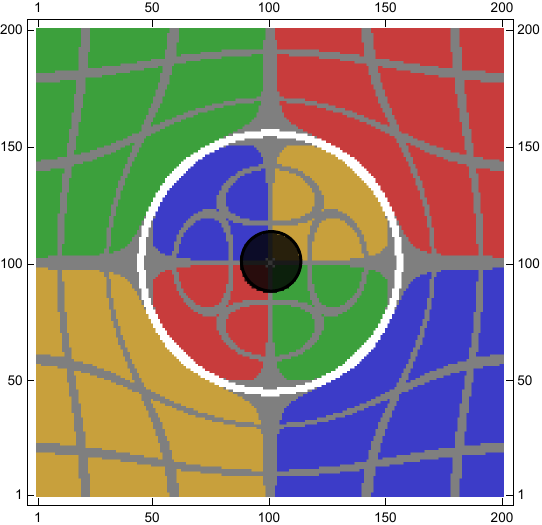}
	}
	
	\vspace{2mm}
	
	\caption{\label{fig5} Optical images and Einstein rings corresponding to different boson stars under a spherical light source, with $\xi=0.05$, $\alpha_{\text{fov}}=45^\circ$, $r_{\text{obs}}=50m$ ($m$ denotes the mass of the boson star), and the inclination angle of the observer is $\theta=45^\circ$.}
\end{figure*}

\begin{figure*}[htbp]
	\centering
	
	% Row 1
	\subfigure[\scriptsize $\psi_0=0.2,\theta=0^\circ$]{%
		\includegraphics[width=0.24\textwidth]{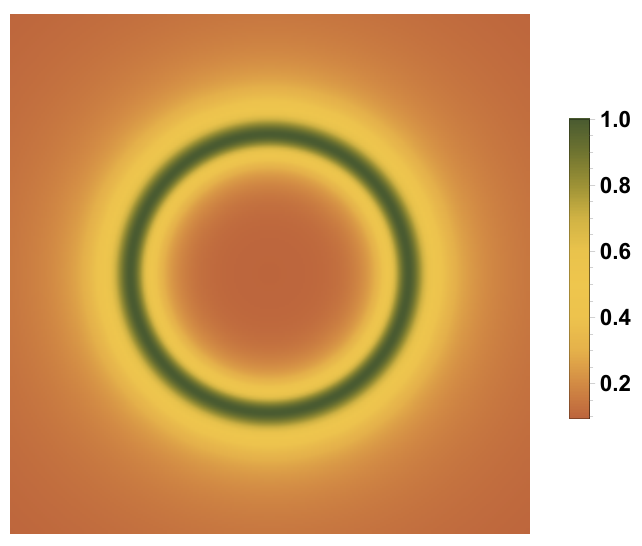}}
	\hfill
	\subfigure[\scriptsize $\psi_0=0.2,\theta=30^\circ$]{%
		\includegraphics[width=0.24\textwidth]{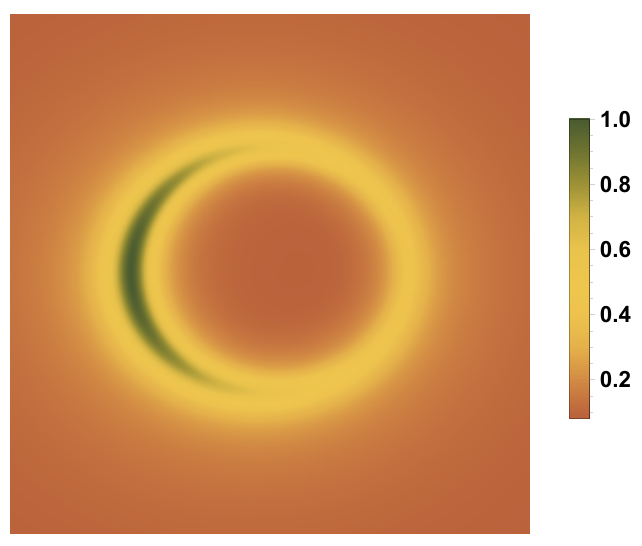}}
	\hfill
	\subfigure[\scriptsize $\psi_0=0.2,\theta=60^\circ$]{%
		\includegraphics[width=0.24\textwidth]{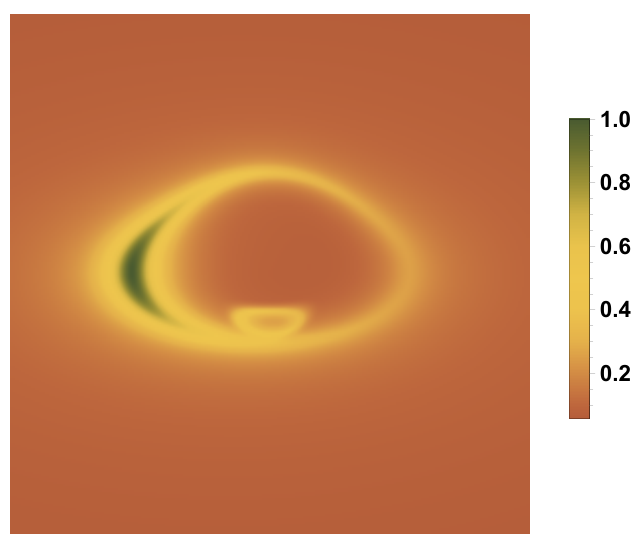}}
	\hfill
	\subfigure[\scriptsize $\psi_0=0.2,\theta=80^\circ$]{%
		\includegraphics[width=0.24\textwidth]{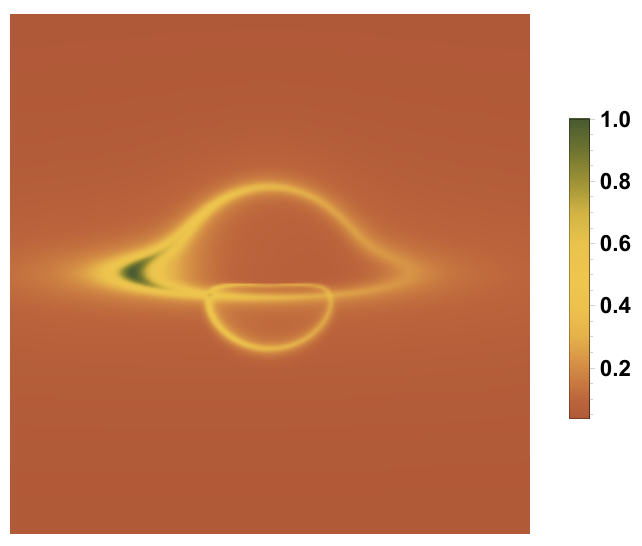}}
	
	\vspace{2mm}
	
	% Row 2
	\subfigure[\scriptsize $\psi_0=0.16,\theta=0^\circ$]{%
		\includegraphics[width=0.24\textwidth]{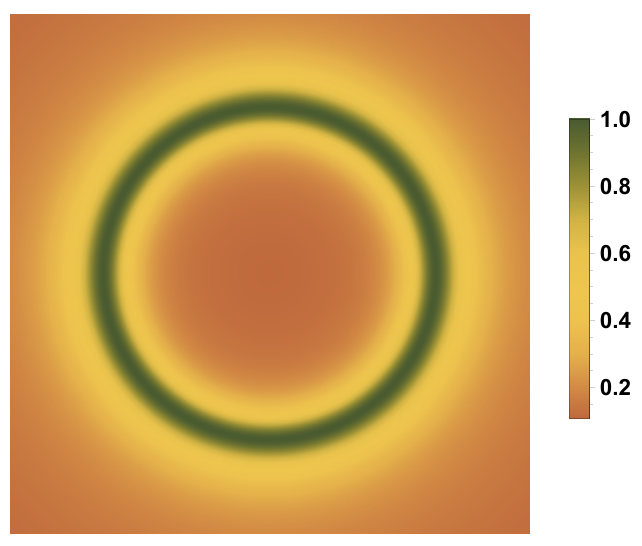}}
	\hfill
	\subfigure[\scriptsize $\psi_0=0.16,\theta=30^\circ$]{%
		\includegraphics[width=0.24\textwidth]{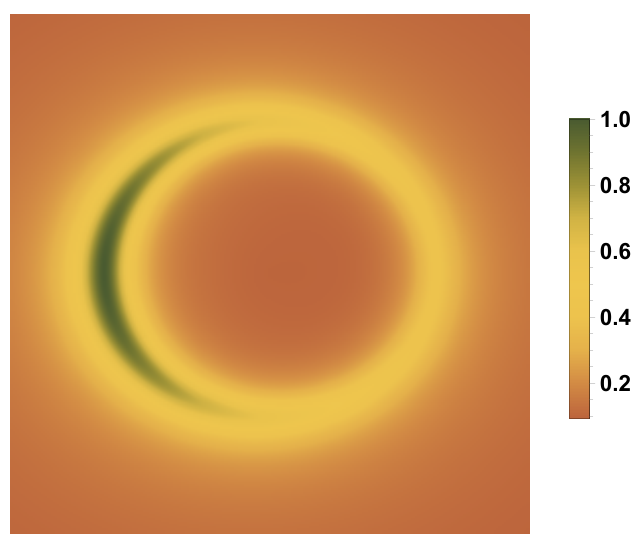}}
	\hfill
	\subfigure[\scriptsize $\psi_0=0.16,\theta=60^\circ$]{%
		\includegraphics[width=0.24\textwidth]{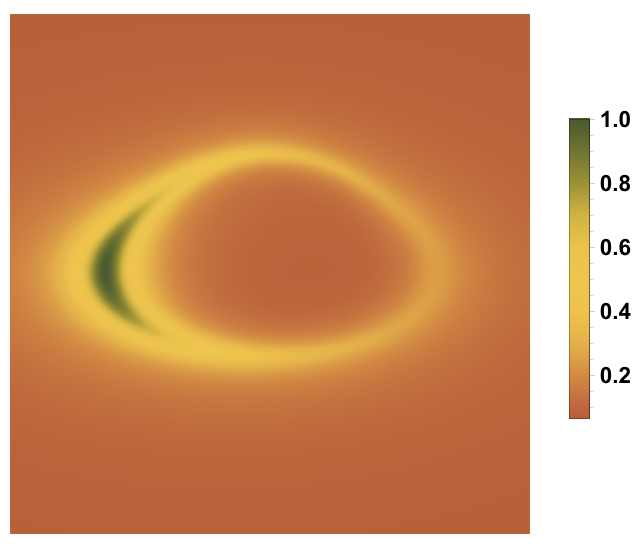}}
	\hfill
	\subfigure[\scriptsize $\psi_0=0.16,\theta=80^\circ$]{%
		\includegraphics[width=0.24\textwidth]{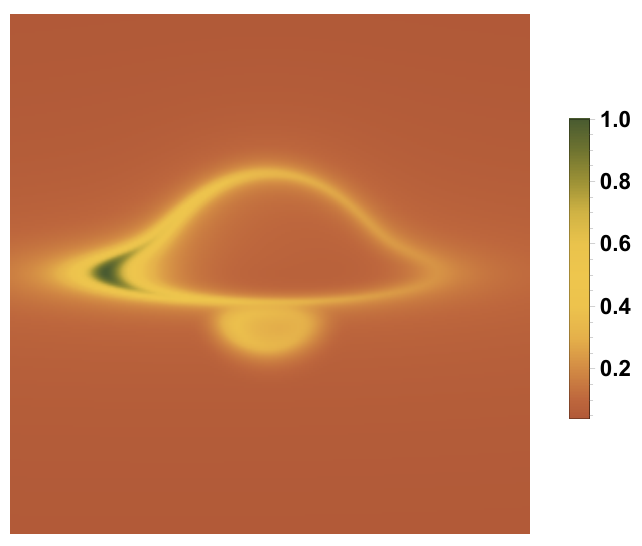}}
	
	\vspace{2mm}
	
	% Row 3
	\subfigure[\scriptsize $\psi_0=0.13,\theta=0^\circ$]{%
		\includegraphics[width=0.24\textwidth]{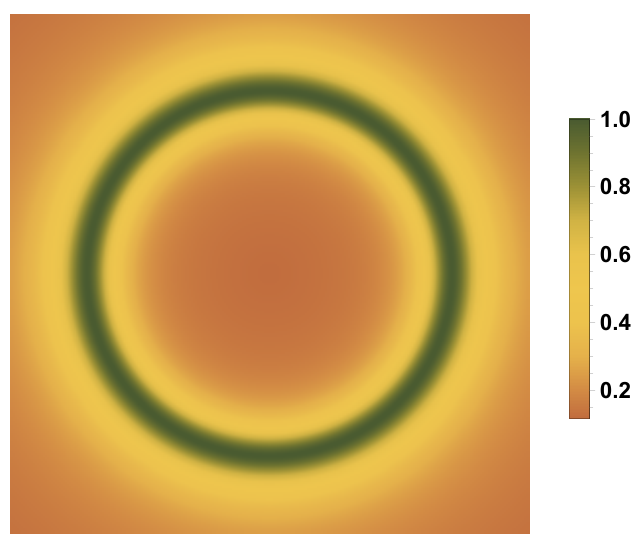}}
	\hfill
	\subfigure[\scriptsize $\psi_0=0.13,\theta=30^\circ$]{%
		\includegraphics[width=0.24\textwidth]{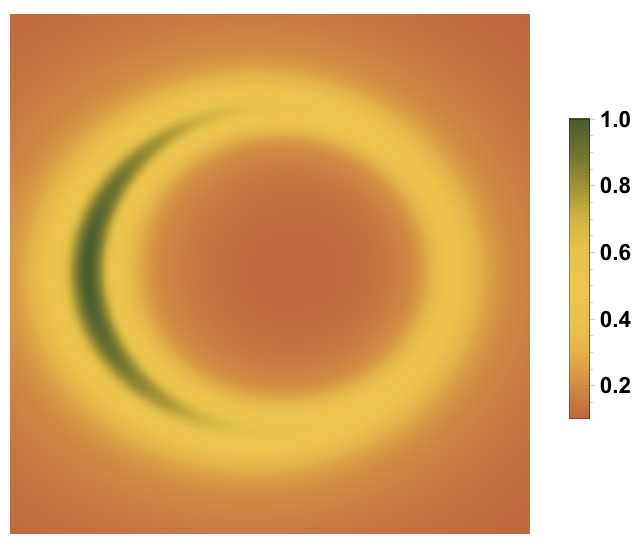}}
	\hfill
	\subfigure[\scriptsize $\psi_0=0.13,\theta=60^\circ$]{%
		\includegraphics[width=0.24\textwidth]{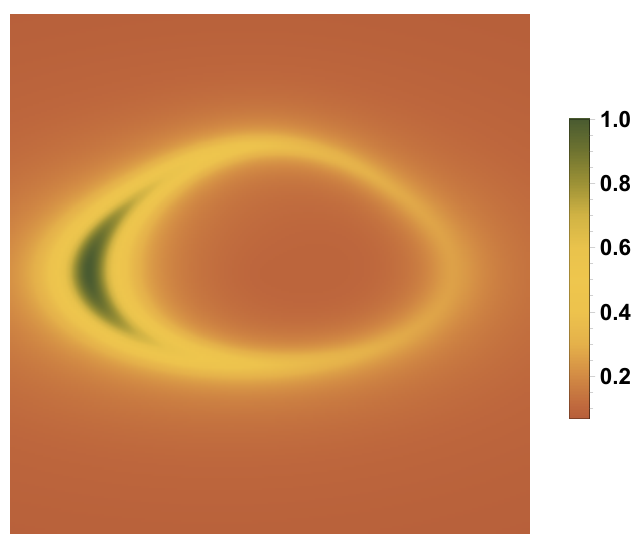}}
	\hfill
	\subfigure[\scriptsize $\psi_0=0.13,\theta=80^\circ$]{%
		\includegraphics[width=0.24\textwidth]{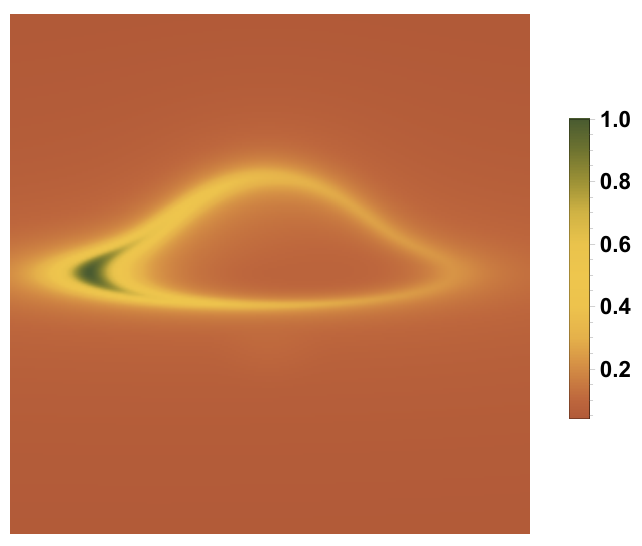}}
	
	\vspace{2mm}
	
	% Row 4
	\subfigure[\scriptsize $\psi_0=0.09,\theta=0^\circ$]{%
		\includegraphics[width=0.24\textwidth]{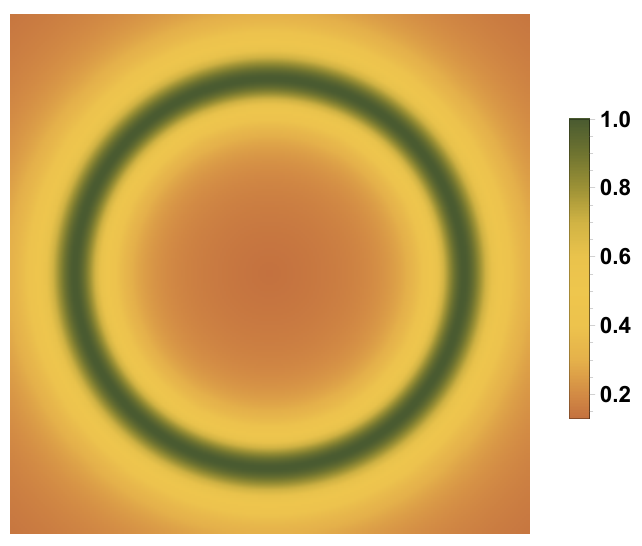}}
	\hfill
	\subfigure[\scriptsize $\psi_0=0.09,\theta=30^\circ$]{%
		\includegraphics[width=0.24\textwidth]{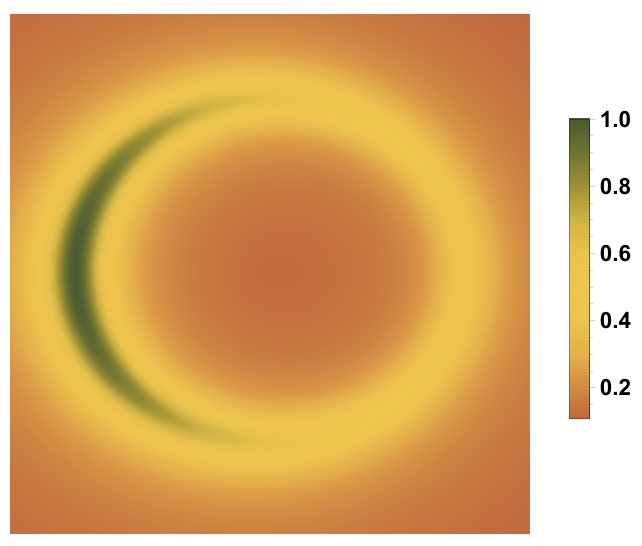}}
	\hfill
	\subfigure[\scriptsize $\psi_0=0.09,\theta=60^\circ$]{%
		\includegraphics[width=0.24\textwidth]{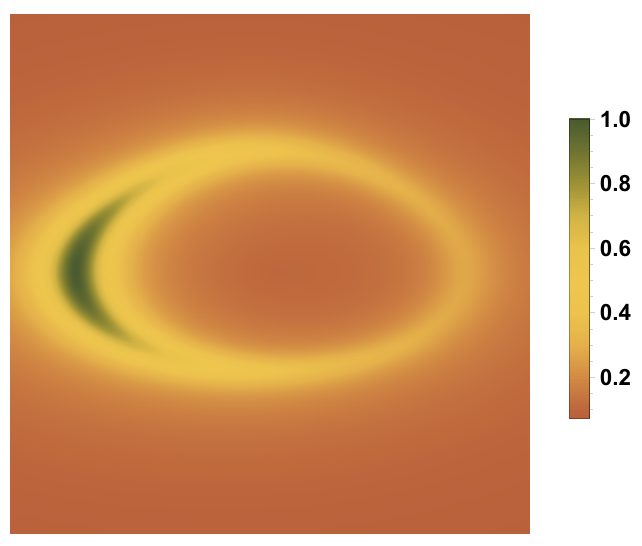}}
	\hfill
	\subfigure[\scriptsize $\psi_0=0.09,\theta=80^\circ$]{%
		\includegraphics[width=0.24\textwidth]{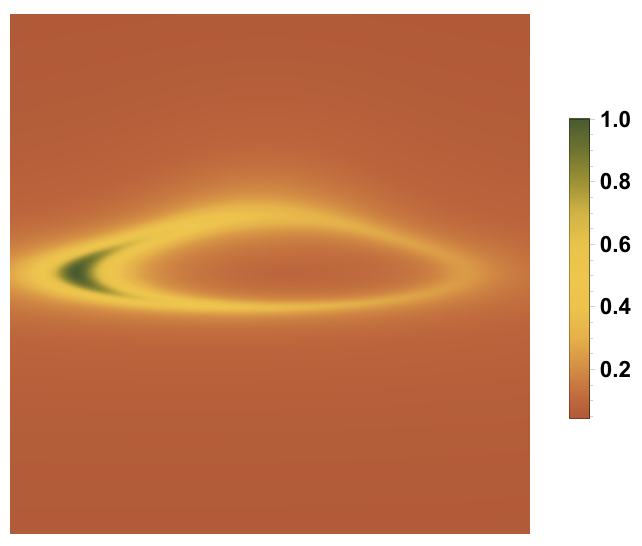}}
	
	\vspace{2mm}
	
	\caption{\label{fig6} Optical images corresponding to different boson stars under a thin accretion disk, with $\xi=0.05$ and $\alpha_{\text{fov}}=3.5^\circ$. Columns from left to right correspond to models with inclination angle $\theta = 0^\circ, 30^\circ, 60^\circ, 80^\circ$, and rows from top to bottom correspond to $\psi_0 = 0.2, 0.16, 0.13, 0.09$.}
\end{figure*}

The optical images of various boson stars illuminated by a spherical light source are shown in Figure~\ref{fig5}. In each image, the celestial sphere is divided into four quadrants, marked in red, green, yellow, and blue, with coordinate lines drawn in gray at $10^\circ$ intervals. The boson star is positioned at the center of the celestial sphere, and the observer is placed at a distance $r_{\text{obs}} = 50m$, defining the radius of the sphere. To investigate strong gravitational lensing and the formation of the Einstein ring, a white reference light source is placed diametrically opposite the observer. The central black transparent region in each image corresponds to the boson star, while the white circular structure in the outer region indicates the Einstein ring.

Several key phenomena are evident from the figures. Since the boson star lacks an event horizon, light entering the boson star is not fully absorbed but is instead partially retained. Similar to a black hole, when the light source passes near the boson star, an Einstein ring forms due to gravitational lensing. Because the boson star is static and does not rotate, its central black region appears as a perfect circle, resembling a non-rotating black hole. Additionally, there is no frame-dragging effect on the background celestial sphere, as the boson star lacks angular momentum. Furthermore, we investigated the effects of the initial scalar field value $\psi_0$ on the optical image. The results show that $\psi_0$ plays a significant role. As $\psi_0$ decreases, both the size of the boson star and the Einstein ring shrink, and the light trajectories inside the Einstein ring change. This highlights the importance of the scalar field's initial conditions in determining the optical characteristics of the boson star.

The optical images of various boson stars with a thin accretion disk are presented in Figure~\ref{fig6}. In these images, the field of view angle is fixed at $\alpha_{\text{fov}} = 3.5^\circ$, the gravitational coupling parameter is set to $\xi = 0.05$, and the observer is located at a distance of $r_{\text{obs}} = 200$. The background is shown in orange, corresponding to zero light intensity, while colors ranging from yellow to green represent increasing intensity, with the boson star positioned at the center of each image.

Several interesting features are observed in these images. When the observation angle $\theta$ is small (e.g., $0^\circ$ and $30^\circ$ in the first and second columns), the optical image shows only the direct image of the boson star. In this case, the scalar field $\psi_0$ primarily influences the size of the direct image without altering its overall shape. As $\psi_0$ decreases, the direct image becomes progressively larger. However, as the observation angle $\theta$ increases (e.g., $60^\circ$ and $80^\circ$ in the third and fourth columns), the effect of $\psi_0$ on the optical image becomes more pronounced, affecting both the size and the shape of the direct image. In these cases, lensed images of light appear in images (c), (d), and (h). For instance, in image (d), the upper hat-shaped optical image represents the direct image ($n = 1$), where light passes through the equatorial plane for the first time, while the lower D-shaped image represents the lensed image ($n = 2$), where light passes through the equatorial plane for the second time.

At $\theta = 0^\circ$, the boson star's optical image appears as a symmetric ring, as the observer’s line of sight is perpendicular to the equatorial plane and aligned with the symmetry axis of both the accretion disk and the boson star. In this case, the observed redshift is primarily attributed to gravitational redshift. As the angle $\theta$ increases, Doppler redshift becomes more significant. A detailed analysis reveals that when $\theta \neq 0$, light approaching the observer is blueshifted, increasing photon energy and brightening the corresponding region in the optical image. Conversely, light receding from the observer is redshifted, reducing photon energy and causing a dimming effect. The images at $\theta = 30^\circ$, $60^\circ$, and $80^\circ$ (second, third, and fourth columns) show that the left side of the image is blueshifted, while the right side is redshifted.

\begin{figure*}[htbp]
	\centering
	
	% Row 1
	\subfigure[\scriptsize $\psi_0=0.2,\theta=0^\circ$]{%
		\includegraphics[width=0.24\textwidth]{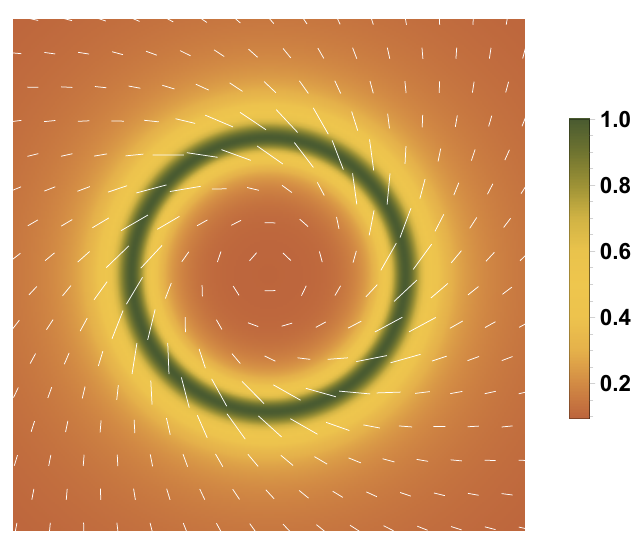}}
	\hfill
	\subfigure[\scriptsize $\psi_0=0.2,\theta=30^\circ$]{%
		\includegraphics[width=0.24\textwidth]{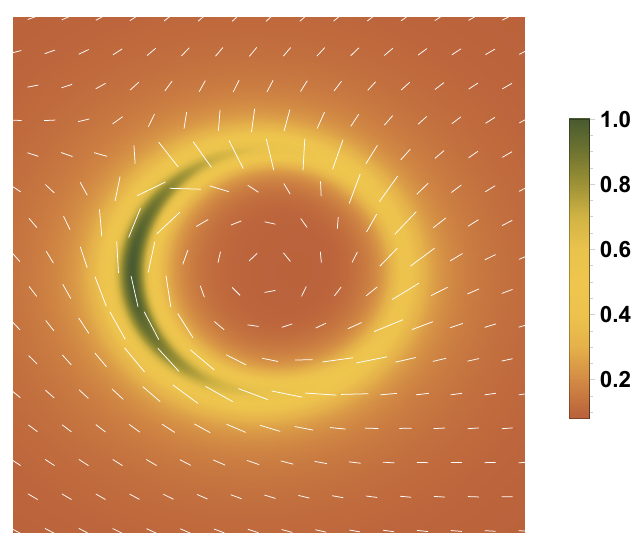}\label{fig7b}}
	\hfill
	\subfigure[\scriptsize $\psi_0=0.2,\theta=60^\circ$]{%
		\includegraphics[width=0.24\textwidth]{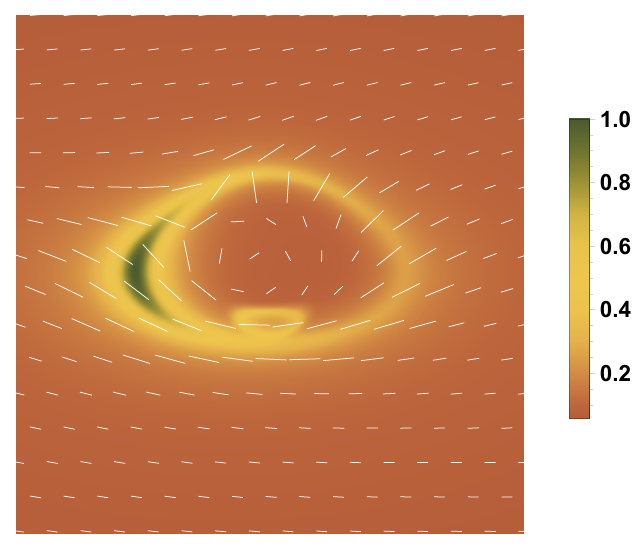}}
	\hfill
	\subfigure[\scriptsize $\psi_0=0.2,\theta=80^\circ$]{%
		\includegraphics[width=0.24\textwidth]{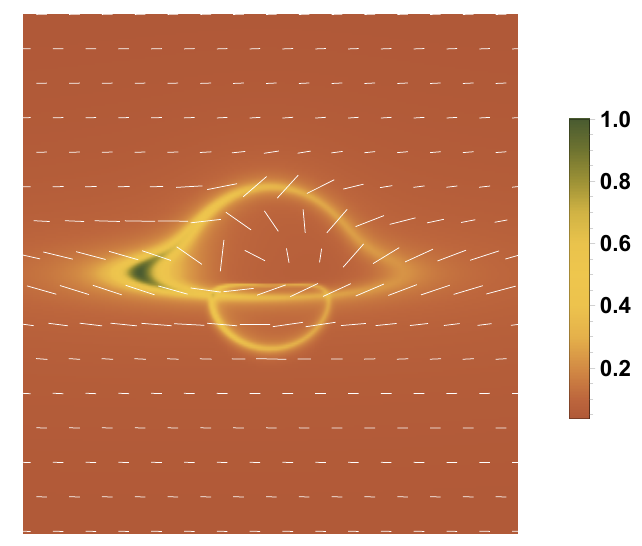}}
	
	\vspace{2mm}
	
	% Row 2
	\subfigure[\scriptsize $\psi_0=0.16,\theta=0^\circ$]{%
		\includegraphics[width=0.24\textwidth]{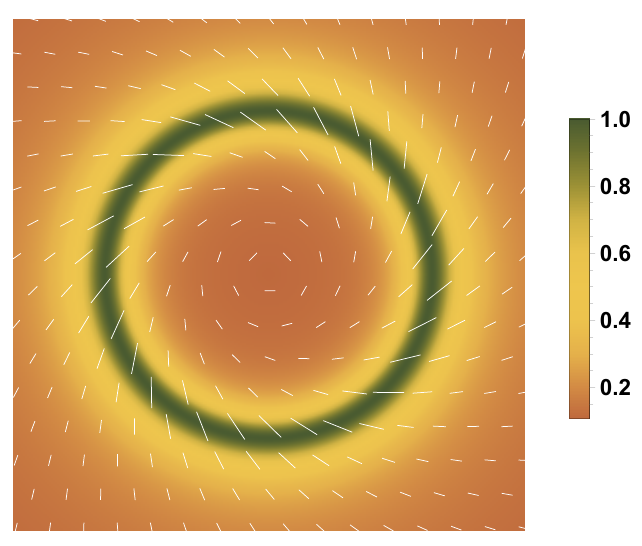}}
	\hfill
	\subfigure[\scriptsize $\psi_0=0.16,\theta=30^\circ$]{%
		\includegraphics[width=0.24\textwidth]{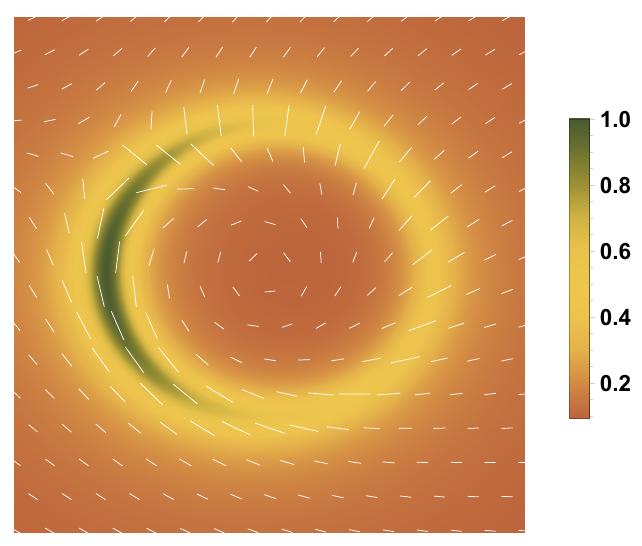}}
	\hfill
	\subfigure[\scriptsize $\psi_0=0.16,\theta=60^\circ$]{%
		\includegraphics[width=0.24\textwidth]{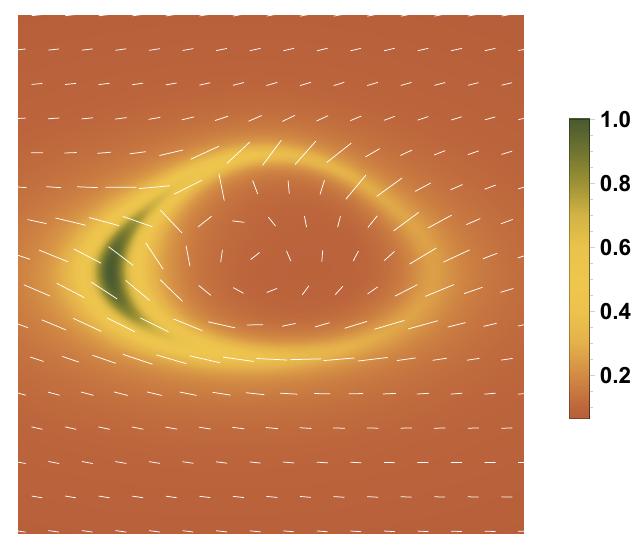}}
	\hfill
	\subfigure[\scriptsize $\psi_0=0.16,\theta=80^\circ$]{%
		\includegraphics[width=0.24\textwidth]{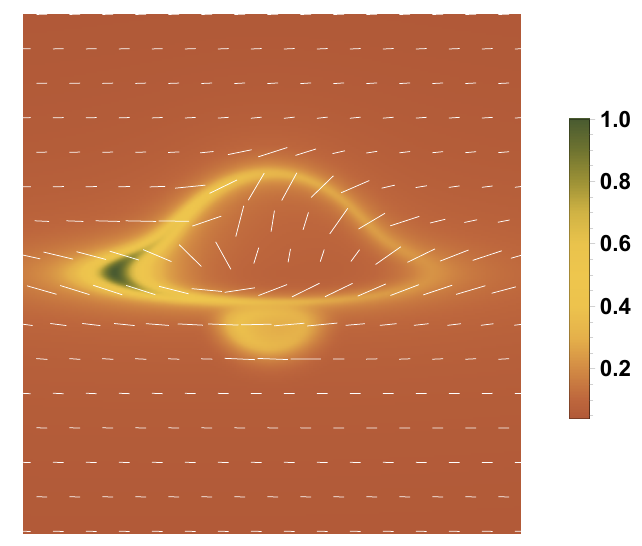}}
	
	\vspace{2mm}
	
	% Row 3
	\subfigure[\scriptsize $\psi_0=0.13,\theta=0^\circ$]{%
		\includegraphics[width=0.24\textwidth]{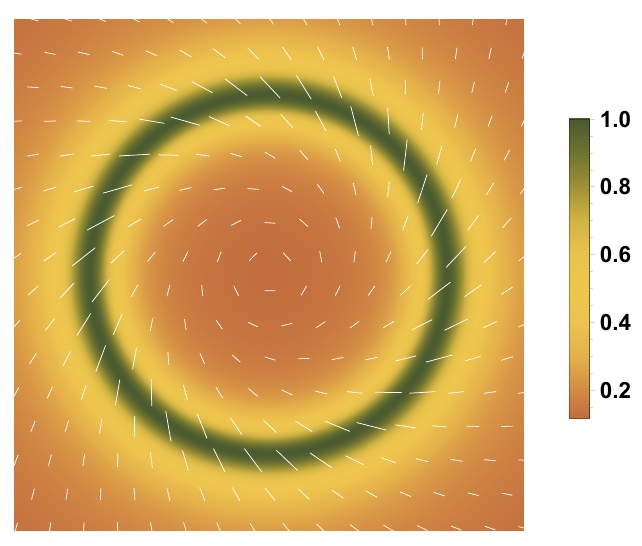}}
	\hfill
	\subfigure[\scriptsize $\psi_0=0.13,\theta=30^\circ$]{%
		\includegraphics[width=0.24\textwidth]{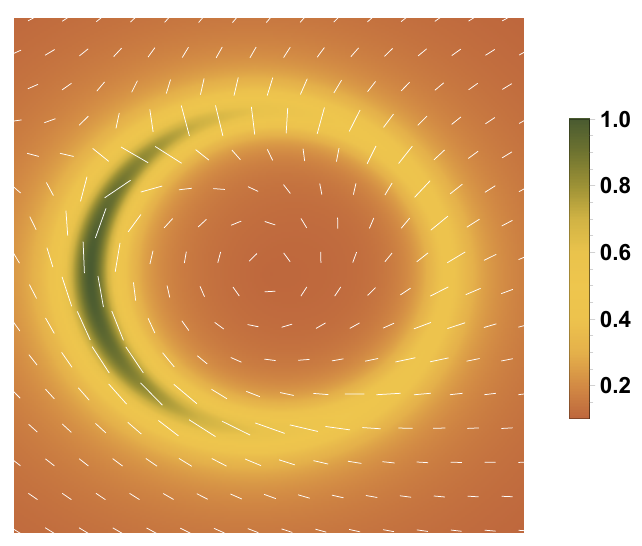}}
	\hfill
	\subfigure[\scriptsize $\psi_0=0.13,\theta=60^\circ$]{%
		\includegraphics[width=0.24\textwidth]{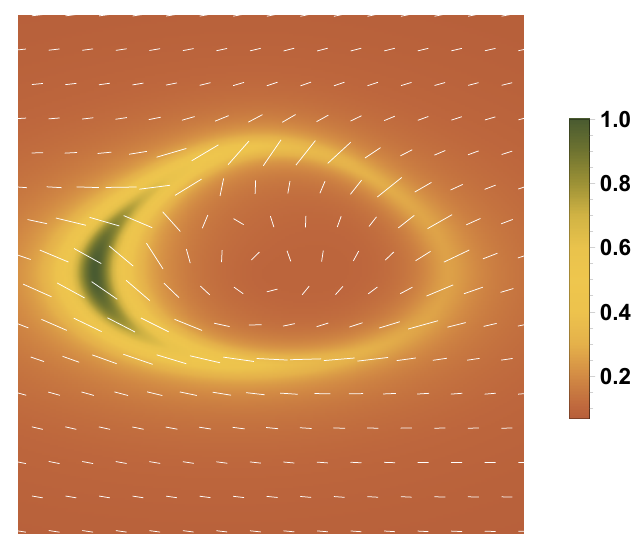}}
	\hfill
	\subfigure[\scriptsize $\psi_0=0.13,\theta=80^\circ$]{%
		\includegraphics[width=0.24\textwidth]{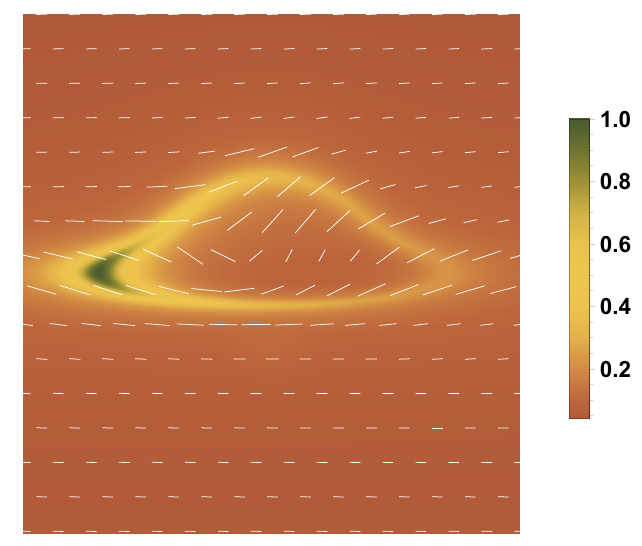}}
	
	\vspace{2mm}
	
	% Row 4
	\subfigure[\scriptsize $\psi_0=0.09,\theta=0^\circ$]{%
		\includegraphics[width=0.24\textwidth]{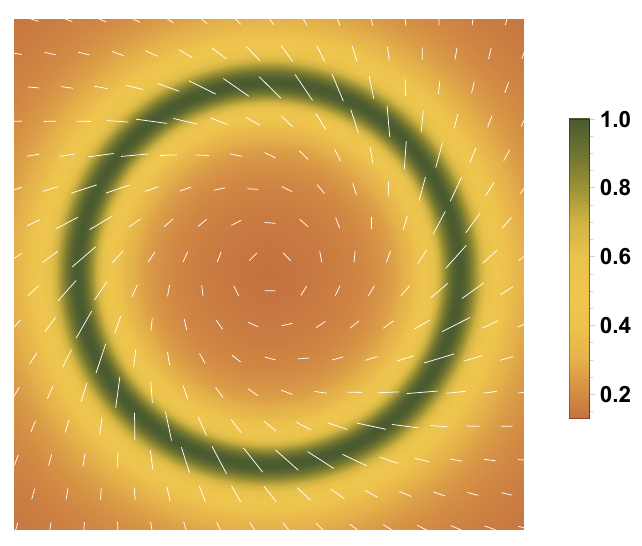}}
	\hfill
	\subfigure[\scriptsize $\psi_0=0.09,\theta=30^\circ$]{%
		\includegraphics[width=0.24\textwidth]{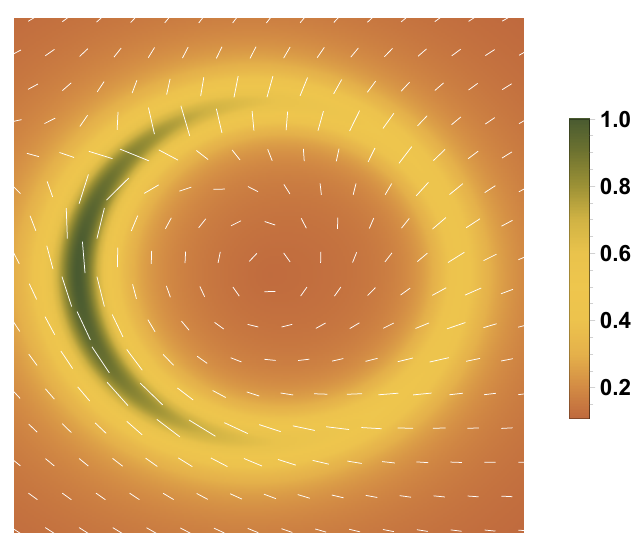}}
	\hfill
	\subfigure[\scriptsize $\psi_0=0.09,\theta=60^\circ$]{%
		\includegraphics[width=0.24\textwidth]{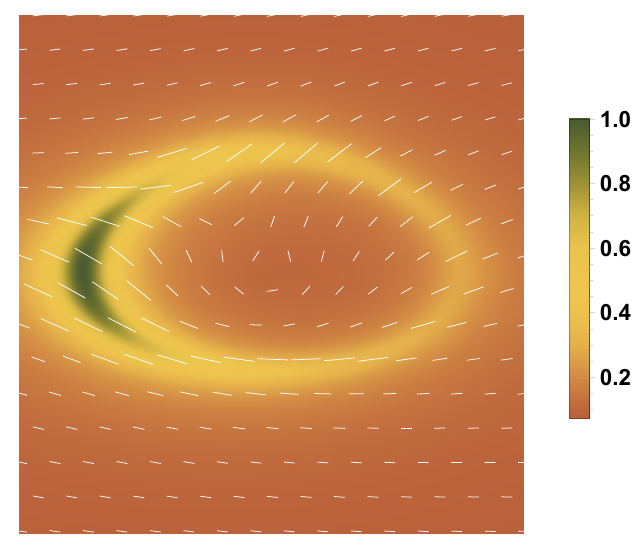}}
	\hfill
	\subfigure[\scriptsize $\psi_0=0.09,\theta=80^\circ$]{%
		\includegraphics[width=0.24\textwidth]{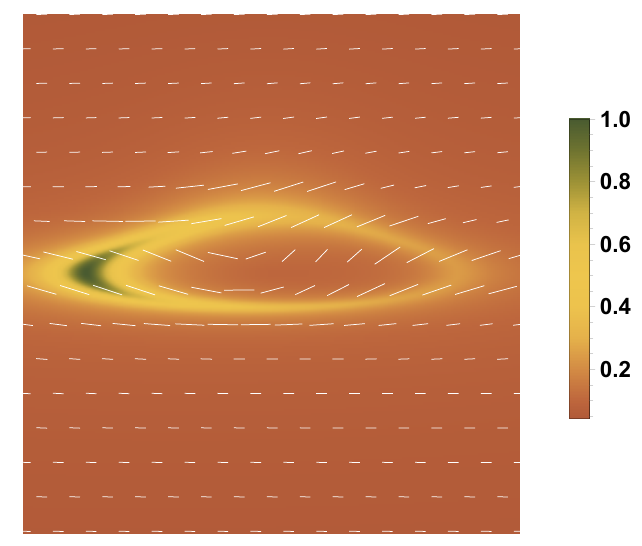}}
	
	\vspace{2mm}
	
	\caption{The effect of the initial scalar field $\psi_0$ on the polarization image. Columns from left to right correspond to models with inclination angle $\theta = 0^\circ, 30^\circ, 60^\circ, 80^\circ$, and rows from top to bottom correspond to $\psi_0 = 0.2, 0.16, 0.13, 0.09$.}
	\label{fig7}
\end{figure*}

Figure~\ref{fig7} shows the effect of $\psi_0$ and $\theta$ on the polarization image. The white lines in the figure represent the polarization vector $\vec{f}$, with the length and direction corresponding to the total linear polarization intensity $\mathsf{P}_o$ and the EVPA $\vartheta$, respectively. For M87*, assuming it is a Schwarzschild black hole, the optimal magnetic field distribution is $\vec{b} = (0.87, 0.5, 0)$, which is used in the simulation. Consistent with previous work, four observer inclination angles $\theta = 0^\circ, 30^\circ, 60^\circ, 80^\circ$ are considered. From the figure, it can be seen that the intensity is positively correlated with $\mathsf{P}_o$, with the polarization intensity significantly enhanced in the brighter regions compared to the darker ones. For example, in Fig.~\ref{fig7b}, the Doppler effect causes the intensity on the left side of the direct image to be greater than that on the right, resulting in a clear left-side polarization intensity distribution. It is worth noting that for polarization images of black holes under the thin accretion disk model, radiation cannot escape outside the event horizon, so theoretically no polarization effects are observable in the inner region. However, for boson stars, the polarization vector can appear across the entire imaging plane, providing an important criterion for distinguishing between these two types of compact objects.

%%%%%%%%%%%%%%%%%%%%%%%%%%%%%%%%%%%%%%

\subsection{Varying gravitational coupling parameter}
\begin{figure}[htbp]
	\centering
	\includegraphics[width=.4\textwidth]{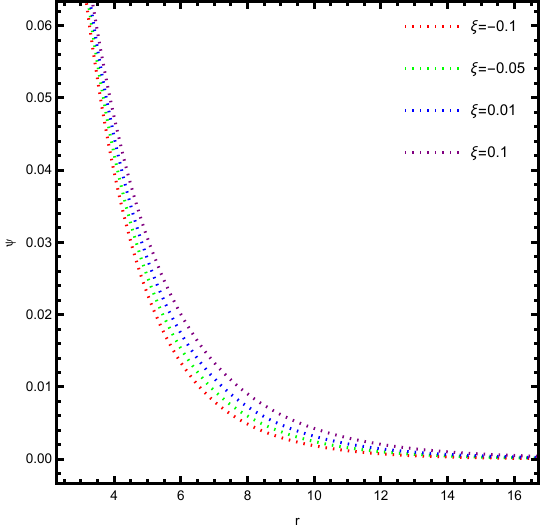}
	\caption{Radial variation of the scalar field for different boson stars, with $\psi_0=0.18$. \label{fig8}}
\end{figure}

\begin{figure}[htbp]
	\centering
	\includegraphics[width=.4\textwidth]{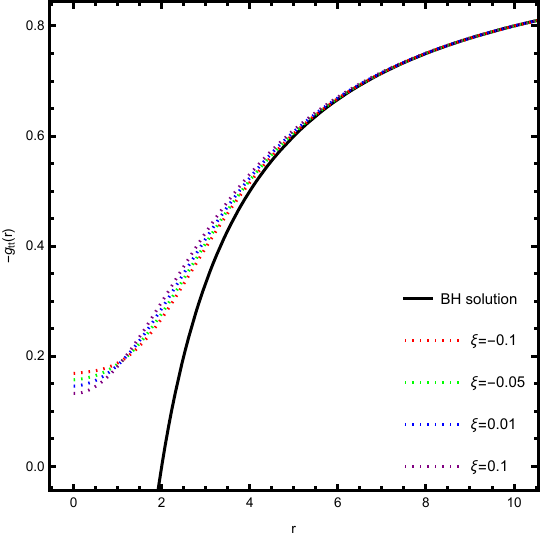}
	\qquad
	\includegraphics[width=.4\textwidth]{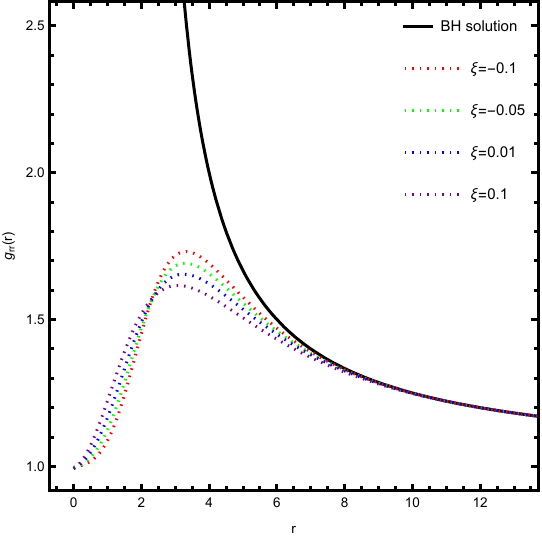}
	\caption{
		Metric components $-g_{tt}$ (left) and $g_{rr}$ (right) for different boson stars and the Schwarzschild black hole, with $\psi_0=0.18$. The mass of the Schwarzschild black hole is set to $1$. The black solid line represents the Schwarzschild black hole, while the dotted lines with different colors represent different boson stars. \label{fig9}}
\end{figure}

In this subsection, we fix the initial scalar field at $\psi_0 = 0.18$ and choose the gravitational coupling parameter $\xi = -0.1, -0.05, 0.01, 0.1$ to investigate the variations in the optical images of boson stars. Figure~\ref{fig8} shows the variation of the scalar field $\psi$ with respect to the radial distance $r$ for different values of $\xi$, where $\psi$ decreases rapidly as $r$ increases and approaches zero as $r \to \infty$. For a fixed $r$, increasing $\xi$ results in higher values of $\psi$. Figure~\ref{fig9} presents the numerical solutions for the metric components $-g_{tt}$ and $g_{rr}$, along with the corresponding components of the Schwarzschild black hole for comparison. The numerical metric components exhibit similar behavior to those in Sec.~\ref{6.1}, remaining finite as $r \to 0$, indicating the absence of an event horizon. For $r > 8M$, the solutions closely match the Schwarzschild metric, confirming the asymptotic flatness of the spacetime.

\begin{figure}[htbp]
	\centering
	\includegraphics[width=.4\textwidth]{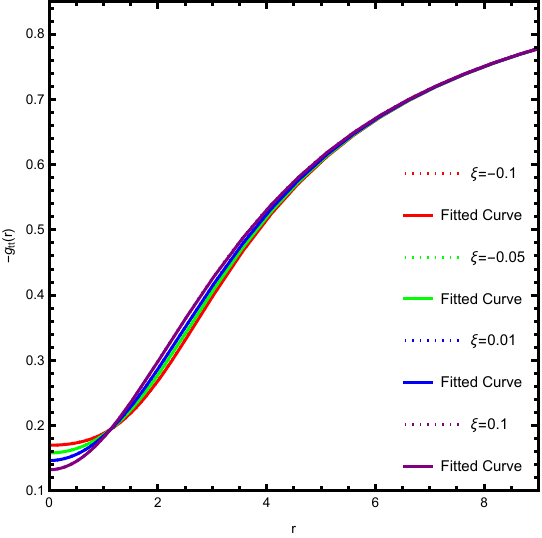}
	\qquad
	\includegraphics[width=.4\textwidth]{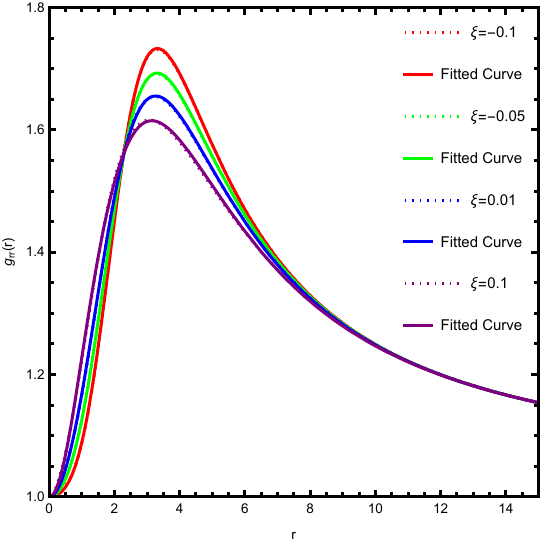}
	\caption{Numerical results and fitting functions of the metric components $-g_{tt}$ (left) and $g_{rr}$ (right) for different boson stars, with $\psi_0=0.18$. The dotted lines represent the numerical results, while the solid lines represent the fitting functions.\label{fig10}}
\end{figure}

\begin{figure}[htbp]
	\centering
	\includegraphics[width=.5\textwidth]{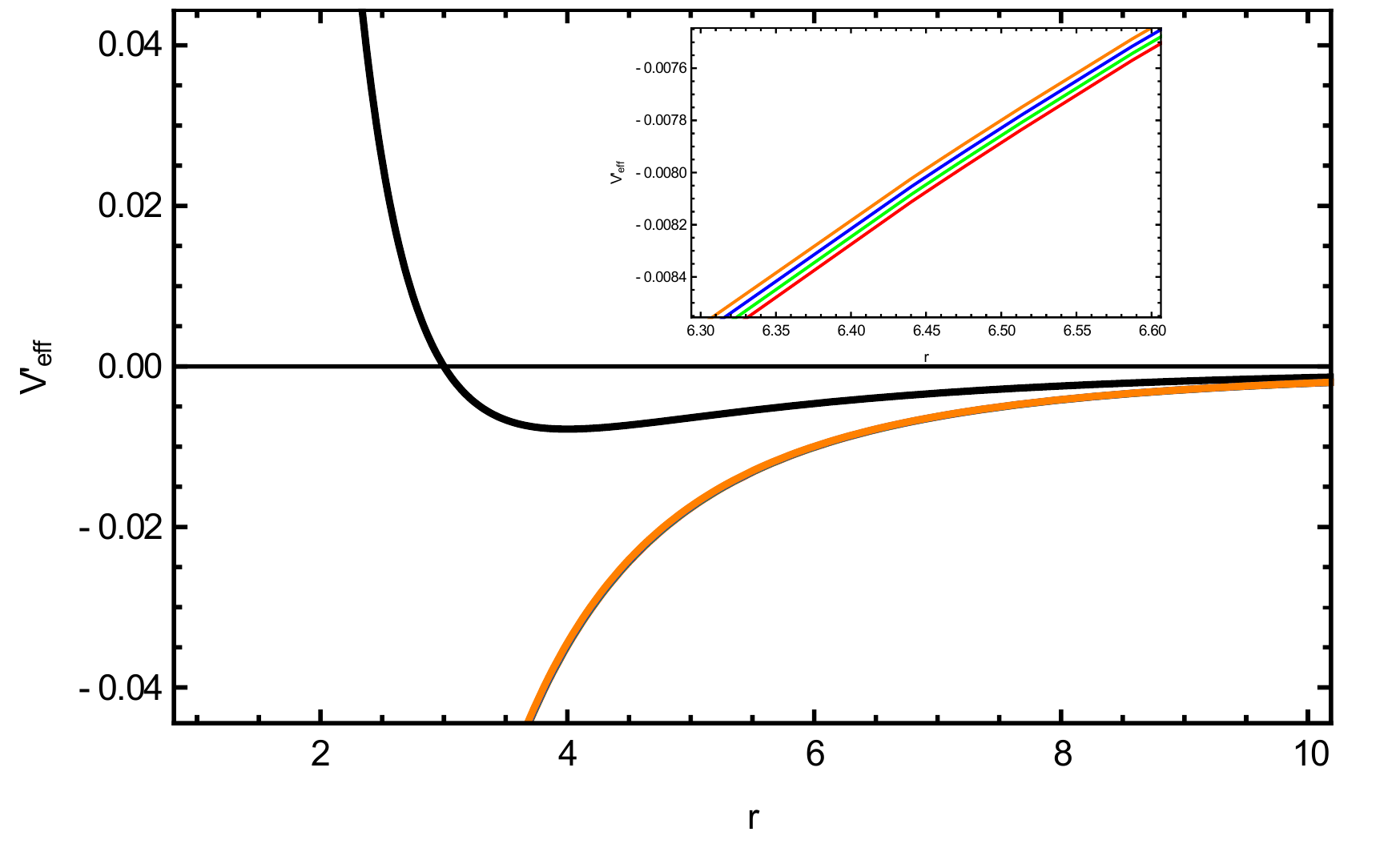}
	\caption{Derivative of the effective potential. The red, blue, green, and orange solid lines correspond to $\xi = -0.1, -0.05, 0.01, 0.1$, with $\psi_0 = 0.18$ fixed. The black solid line represents a Schwarzschild black hole with mass $1$.\label{fig11}}
\end{figure}

The fitting results for the metric components are shown in Figure~\ref{fig10}, where the numerical results are plotted as dotted lines and the fitting functions as solid lines. It is evident that for both $-g_{tt}$ and $g_{rr}$, the numerical results closely match the fitting functions, allowing the fitted metric to be used in subsequent calculations. Tables~\ref{tab3} and~\ref{tab4} summarize the estimated values of the fitting function parameters and the boson star mass $m$ for different values of $\xi$. It is observed that as $\xi$ increases, the mass $m$ decreases. Figure~\ref{fig11} shows the derivative of the effective potential. For all values of $\xi$, the effective potential $V_{\mathrm{eff}}(r)$ does not vanish, indicating the absence of photon rings, as also illustrated in Figure~\ref{fig13}.

\begin{table*}[htbp]
	\centering
	\caption{\label{tab3}Estimated values of the parameters $\alpha_i$ $(i=1,2,\dots,7)$ in the fitting function $g_{tt}$ [see equation~(\ref{gtt})], where $m$ denotes the mass of the boson star and $\psi_0 = 0.18$ is fixed.}
	\begin{tabular}{lccccccccc}
		\toprule
		Type & $\xi$ & $m$ & $\alpha_1$ & $\alpha_2$ & $\alpha_3$ & $\alpha_4$ & $\alpha_5$ & $\alpha_6$ & $\alpha_7$ \\
		\midrule
		BS1 & -0.10 & 0.560 & -0.183 & 0.341 & -0.367 & 0.061  & -0.118 & -0.038 & -0.125 \\
		BS2 & -0.05 & 0.531 & -0.049 & 0.324 & -0.356 & 0.011  & -0.128 & -0.036 & -0.118 \\
		BS3 &  0.01 & 0.502 &  0.150 & 0.287 & -0.331 & -0.054 & -0.136 & -0.028 & -0.099 \\
		BS4 &  0.10 & 0.470 &  0.601 & 0.234 & -0.263 & -0.157 & -0.132 & -0.012 & -0.046 \\
		\bottomrule
	\end{tabular}
\end{table*}

\begin{table*}[htbp]
	\centering
	\caption{\label{tab4}Estimated values of the parameters $\beta_i$ $(i=1,2,\dots,7)$ in the fitting function $g_{rr}$ [see equation~(\ref{grr})], where $m$ denotes the mass of the boson star and $\psi_0 = 0.18$ is fixed.}
	\begin{tabular}{lccccccccc}
		\toprule
		Type & $\xi$ & $m$ & $\beta_1$ & $\beta_2$ & $\beta_3$ & $\beta_4$ & $\beta_5$ & $\beta_6$ & $\beta_7$ \\
		\midrule
		BS1 & -0.10 & 0.560 & -12.309 & -5.858 & 2.986 & -0.085 & 2.184 & 0.071 & -0.011 \\
		BS2 & -0.05 & 0.531 & -17.078 & -6.108 & 4.105 &  0.666 & 2.874 & 0.133 & -0.022 \\
		BS3 &  0.01 & 0.502 & -19.516 & -3.857 & 3.507 &  2.172 & 2.424 & 0.103 & -0.028 \\
		BS4 &  0.10 & 0.470 & -24.078 & -1.971 & 2.314 &  4.192 & 1.786 & 0.038 & -0.023 \\
		\bottomrule
	\end{tabular}
\end{table*}

\begin{figure*}[htbp]
	\centering
	
	\subfigure[\scriptsize $\xi=-0.1,\theta=45^\circ$]{%
		\includegraphics[width=0.22\textwidth]{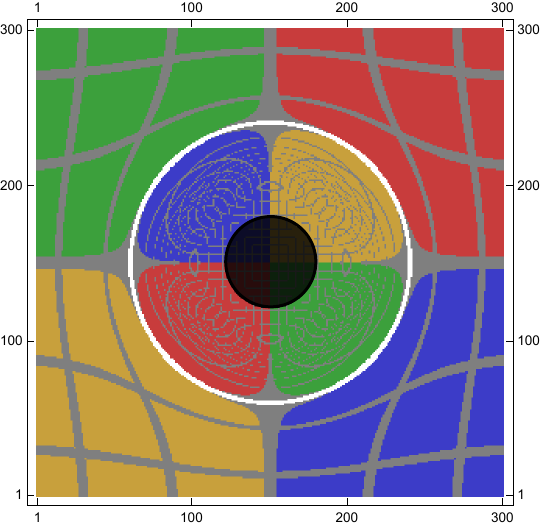}}
	\hfill
	\subfigure[\scriptsize $\xi=-0.05,\theta=45^\circ$]{%
		\includegraphics[width=0.22\textwidth]{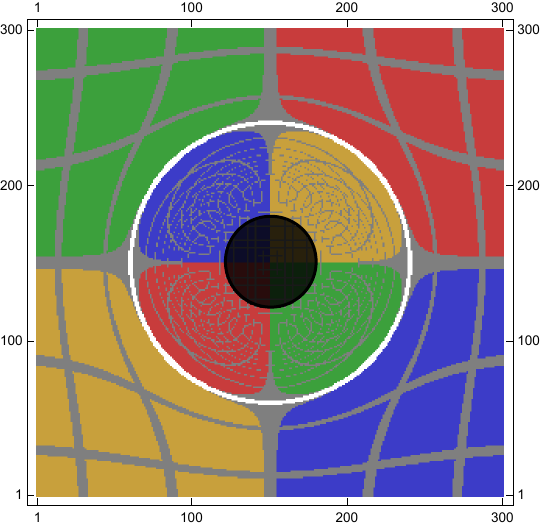}}
	\hfill
	\subfigure[\scriptsize $\xi=0.01,\theta=45^\circ$]{%
		\includegraphics[width=0.22\textwidth]{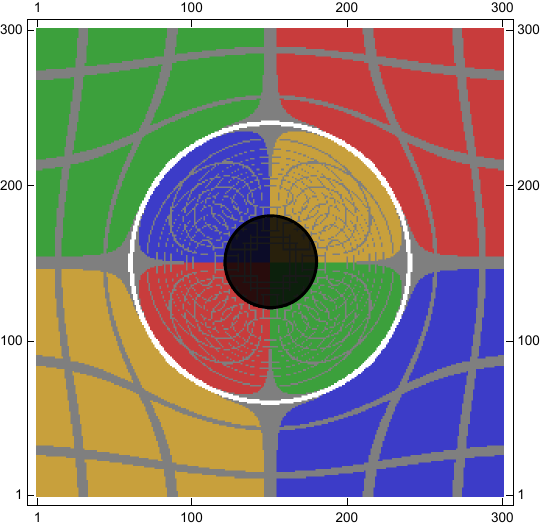}}
	\hfill
	\subfigure[\scriptsize $\xi=0.1,\theta=45^\circ$]{%
		\includegraphics[width=0.22\textwidth]{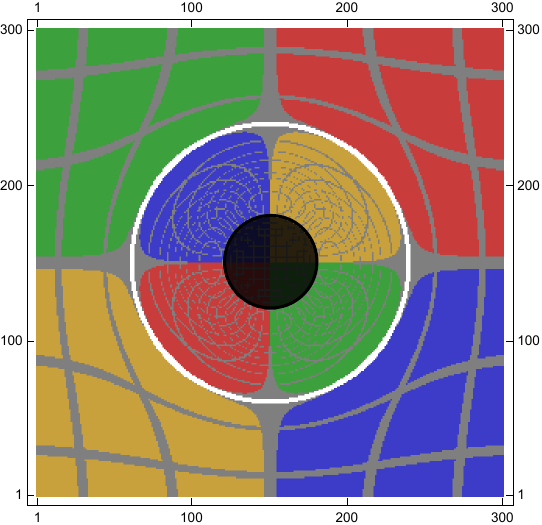}}
	
	\vspace{2mm}
	
	\caption{\label{fig12} Optical images and Einstein rings corresponding to different boson stars under a spherical light source, with $\psi_0 = 0.18$, $\alpha_{\text{fov}}=45^\circ$, $r_{\text{obs}}=50m$ ($m$ denotes the mass of the boson star), and the inclination angle of the observer is $\theta=45^\circ$.}
\end{figure*}

\begin{figure*}[htbp]
	\centering
	
	% Row 1
	\subfigure[\scriptsize $\xi=-0.1,\theta=0^\circ$]{%
		\includegraphics[width=0.24\textwidth]{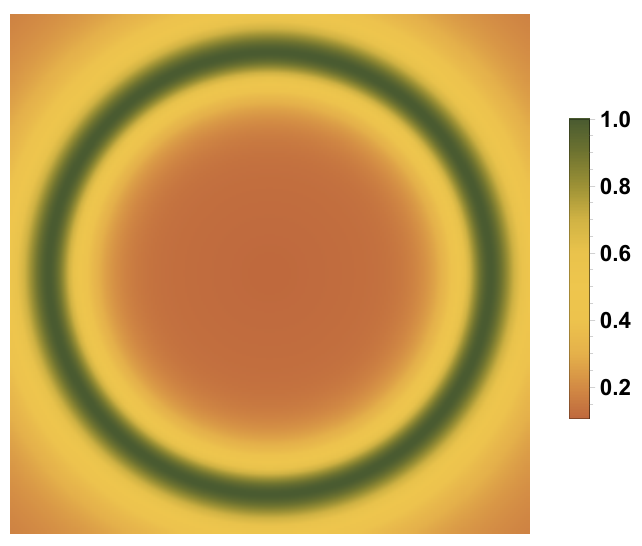}}
	\hfill
	\subfigure[\scriptsize $\xi=-0.1,\theta=30^\circ$]{%
		\includegraphics[width=0.24\textwidth]{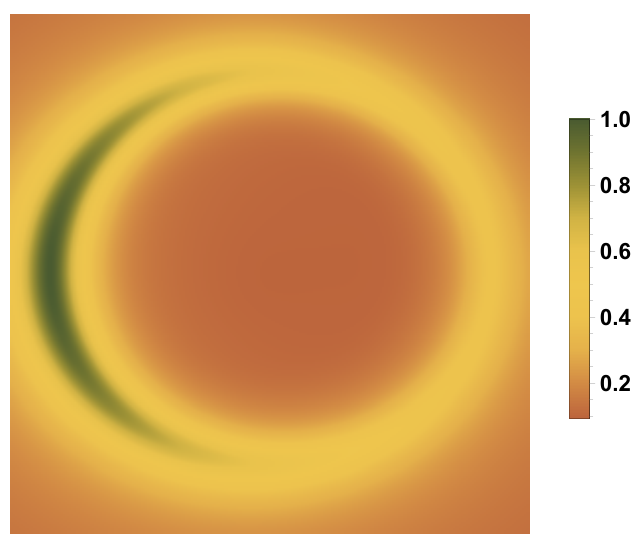}}
	\hfill
	\subfigure[\scriptsize $\xi=-0.1,\theta=60^\circ$]{%
		\includegraphics[width=0.24\textwidth]{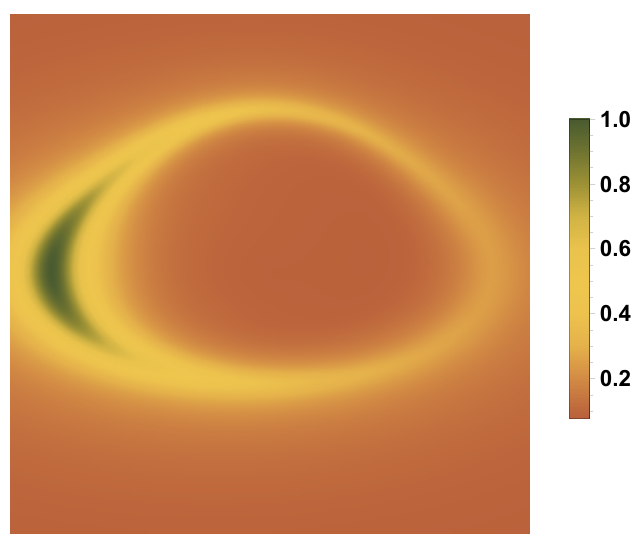}}
	\hfill
	\subfigure[\scriptsize $\xi=-0.1,\theta=86^\circ$]{%
		\includegraphics[width=0.24\textwidth]{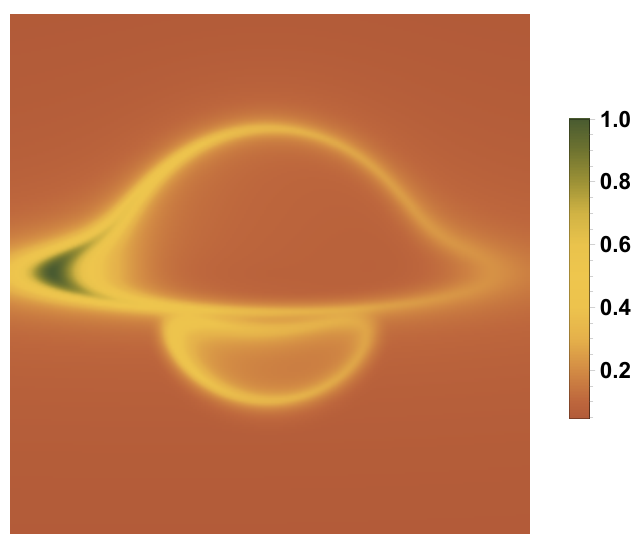}}
	
	\vspace{2mm}
	
	% Row 2
	\subfigure[\scriptsize $\xi=-0.05,\theta=0^\circ$]{%
		\includegraphics[width=0.24\textwidth]{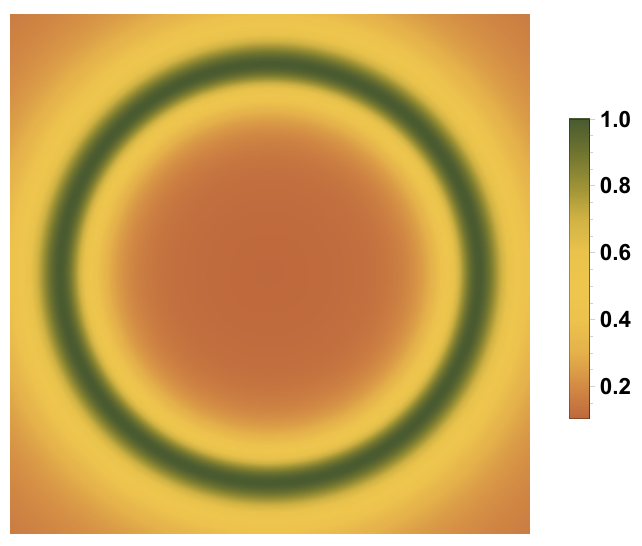}}
	\hfill
	\subfigure[\scriptsize $\xi=-0.05,\theta=30^\circ$]{%
		\includegraphics[width=0.24\textwidth]{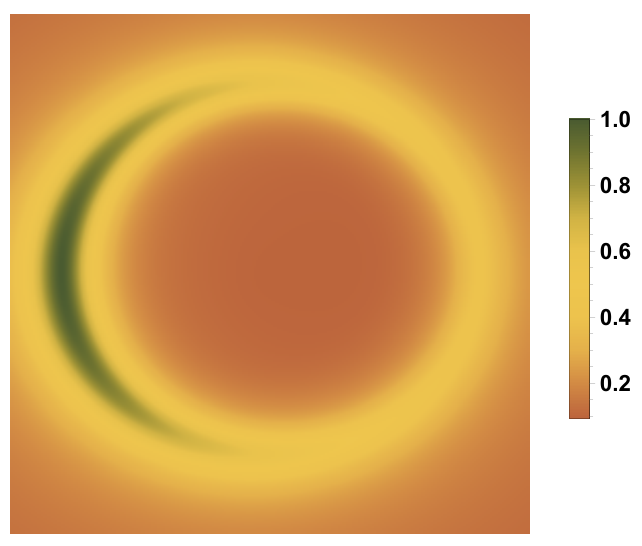}}
	\hfill
	\subfigure[\scriptsize $\xi=-0.05,\theta=60^\circ$]{%
		\includegraphics[width=0.24\textwidth]{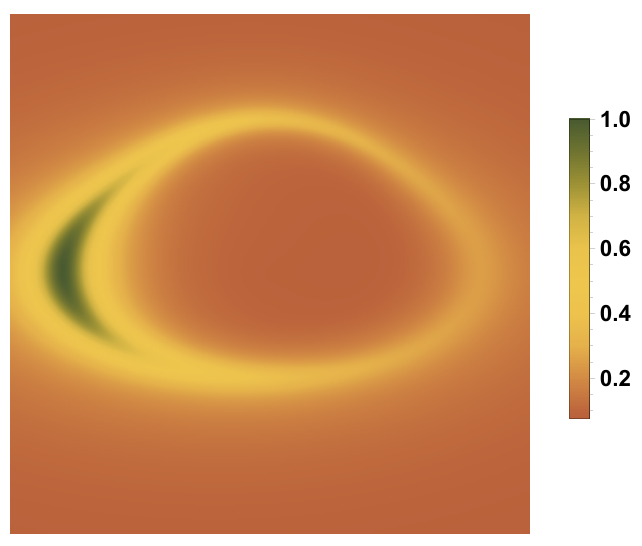}}
	\hfill
	\subfigure[\scriptsize $\xi=-0.05,\theta=86^\circ$]{%
		\includegraphics[width=0.24\textwidth]{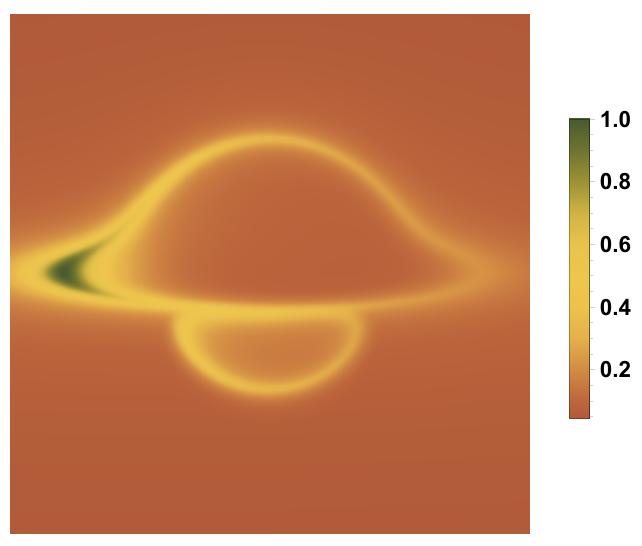}}
	
	\vspace{2mm}
	
	% Row 3
	\subfigure[\scriptsize $\xi=0.01,\theta=0^\circ$]{%
		\includegraphics[width=0.24\textwidth]{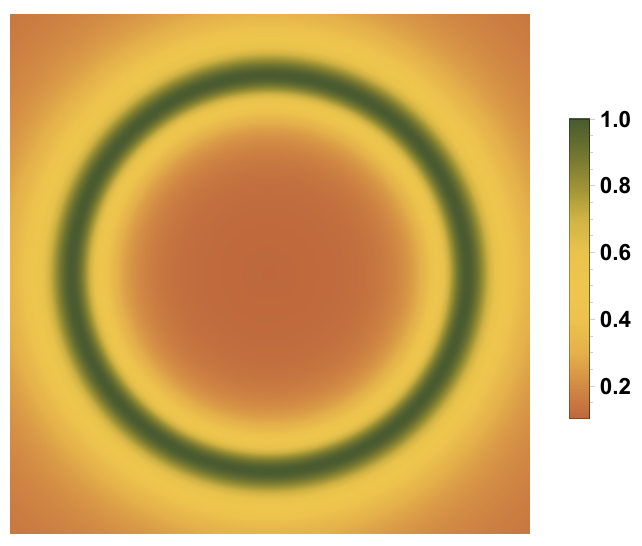}}
	\hfill
	\subfigure[\scriptsize $\xi=0.01,\theta=30^\circ$]{%
		\includegraphics[width=0.24\textwidth]{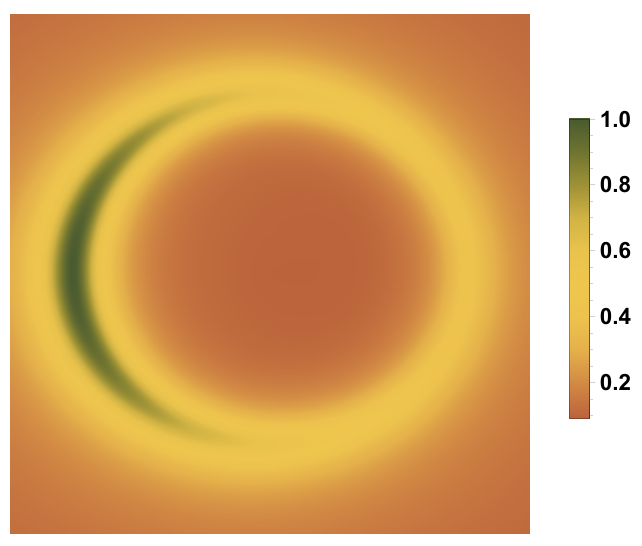}}
	\hfill
	\subfigure[\scriptsize $\xi=0.01,\theta=60^\circ$]{%
		\includegraphics[width=0.24\textwidth]{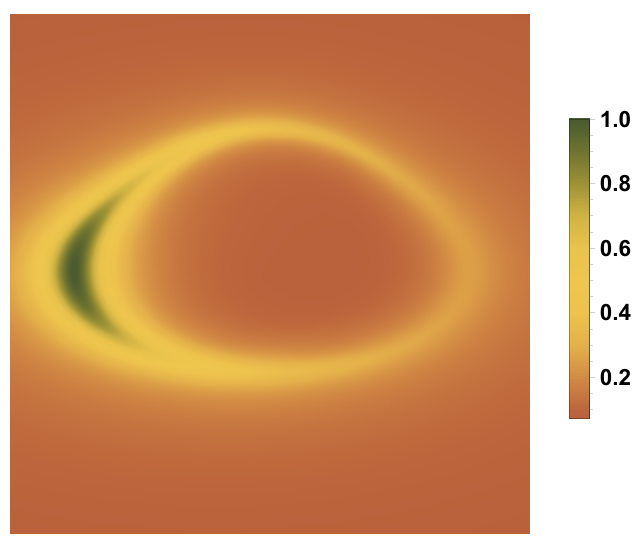}}
	\hfill
	\subfigure[\scriptsize $\xi=0.01,\theta=86^\circ$]{%
		\includegraphics[width=0.24\textwidth]{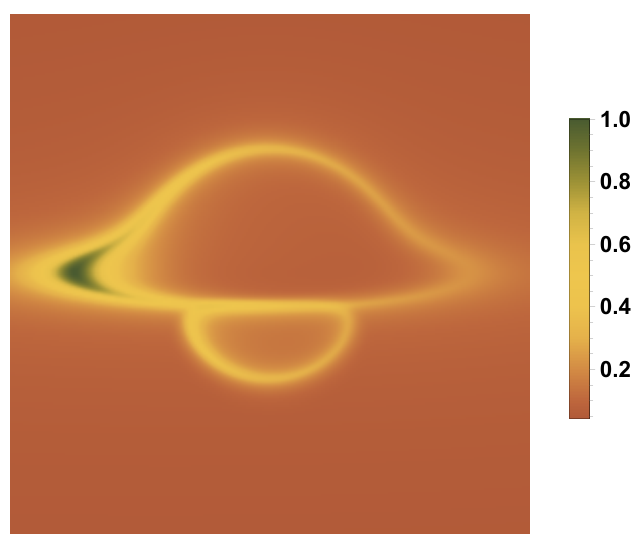}}
	
	\vspace{2mm}
	
	% Row 4
	\subfigure[\scriptsize $\xi=0.1,\theta=0^\circ$]{%
		\includegraphics[width=0.24\textwidth]{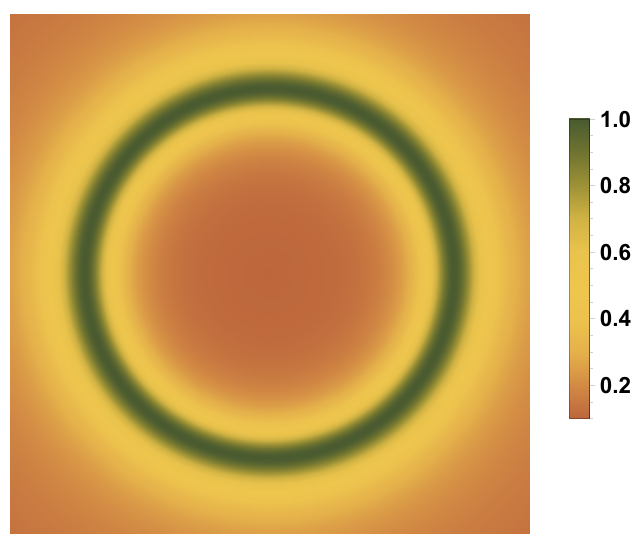}}
	\hfill
	\subfigure[\scriptsize $\xi=0.1,\theta=30^\circ$]{%
		\includegraphics[width=0.24\textwidth]{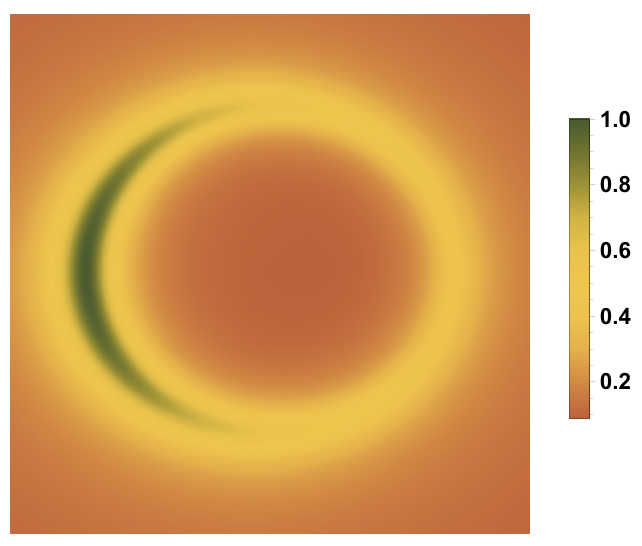}}
	\hfill
	\subfigure[\scriptsize $\xi=0.1,\theta=60^\circ$]{%
		\includegraphics[width=0.24\textwidth]{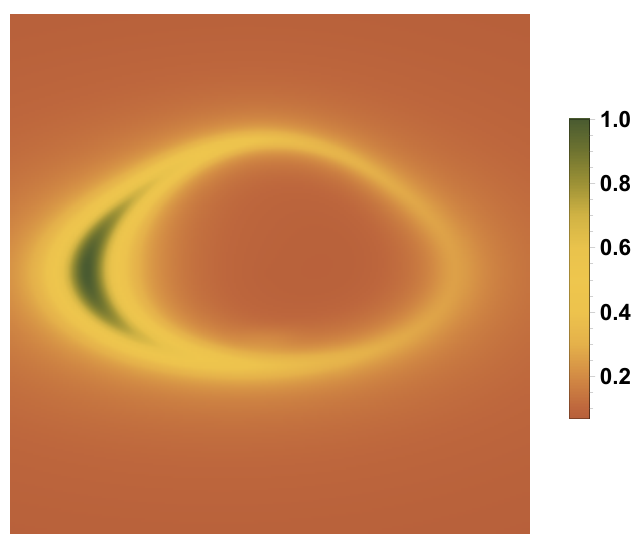}}
	\hfill
	\subfigure[\scriptsize $\xi=0.1,\theta=86^\circ$]{%
		\includegraphics[width=0.24\textwidth]{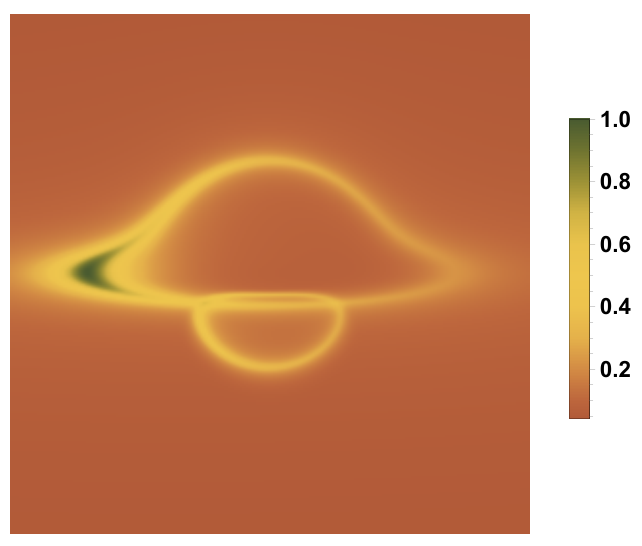}}
	
	\vspace{2mm}
	
	\caption{\label{fig13} Optical images corresponding to different boson stars under a thin accretion disk, with $\psi_0 = 0.18$ and $\alpha_{\text{fov}}=2.8^\circ$. Columns from left to right correspond to models with inclination angle $\theta = 0^\circ, 30^\circ, 60^\circ, 86^\circ$, and rows from top to bottom correspond to $\xi = -0.1, -0.05, 0.01, 0.1$.}
\end{figure*}

The optical images of different boson stars corresponding to a spherical light source are presented in Figure~\ref{fig12}. It is observed that as $\xi$ increases, the sizes of both the boson star and the Einstein ring remain nearly unchanged. However, the light trajectories within the Einstein ring exhibit notable variations. This can be attributed to changes in the gravitational field near the boson star induced by variations in $\xi$, which in turn modify the null geodesics.

%%%%%%%%%%%%%%%%%%%%%%%%%%%%%%%%%%%%%%%%%%%%%%

\begin{figure*}[htbp]
	\centering
	
		% Row 1
	\subfigure[\scriptsize $\xi=-0.1,\theta=0^\circ$]{%
		\includegraphics[width=0.24\textwidth]{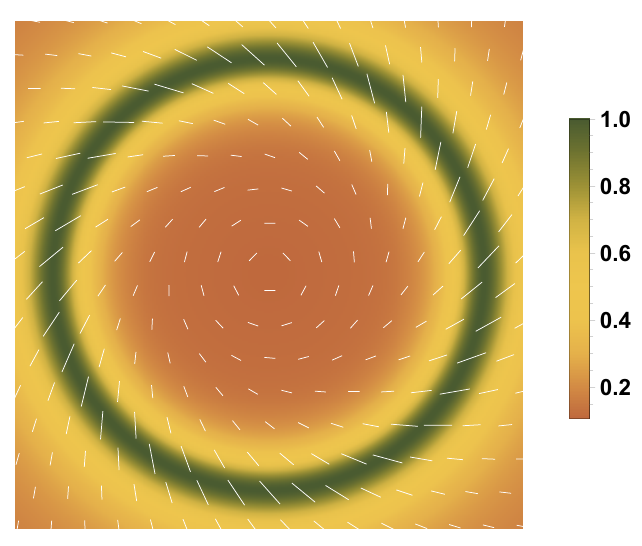}}
	\hfill
	\subfigure[\scriptsize $\xi=-0.1,\theta=30^\circ$]{%
		\includegraphics[width=0.24\textwidth]{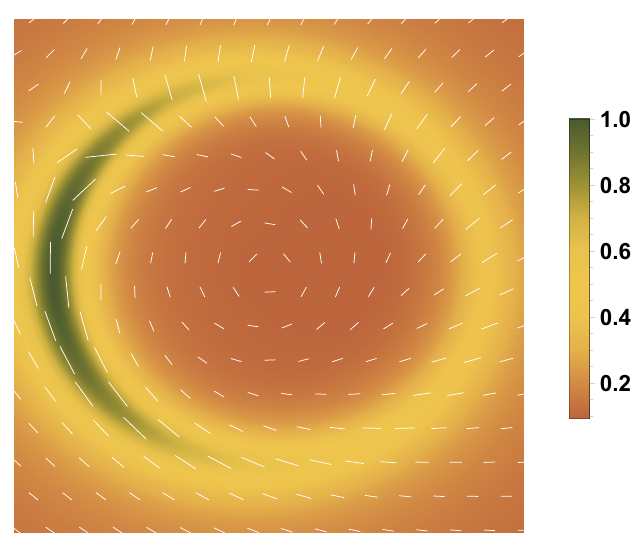}}
	\hfill
	\subfigure[\scriptsize $\xi=-0.1,\theta=60^\circ$]{%
		\includegraphics[width=0.24\textwidth]{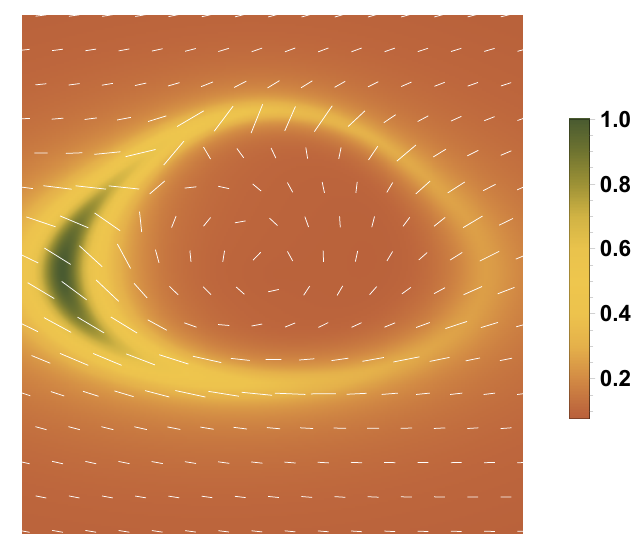}}
	\hfill
	\subfigure[\scriptsize $\xi=-0.1,\theta=86^\circ$]{%
		\includegraphics[width=0.24\textwidth]{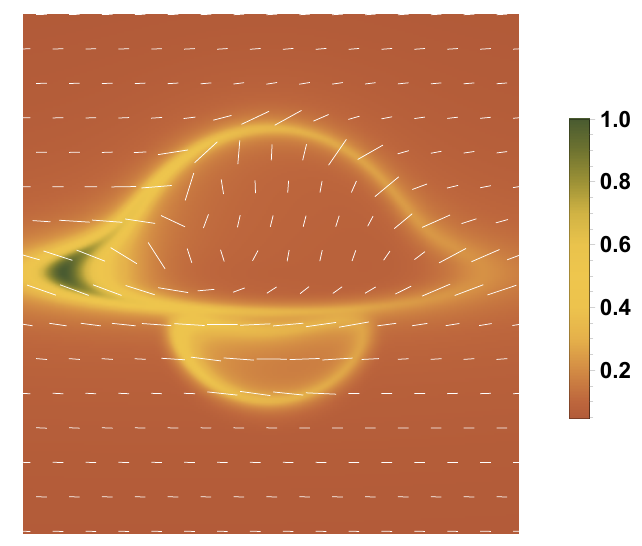}}
	
	\vspace{2mm}
	
	% Row 2
	\subfigure[\scriptsize $\xi=-0.05,\theta=0^\circ$]{%
		\includegraphics[width=0.24\textwidth]{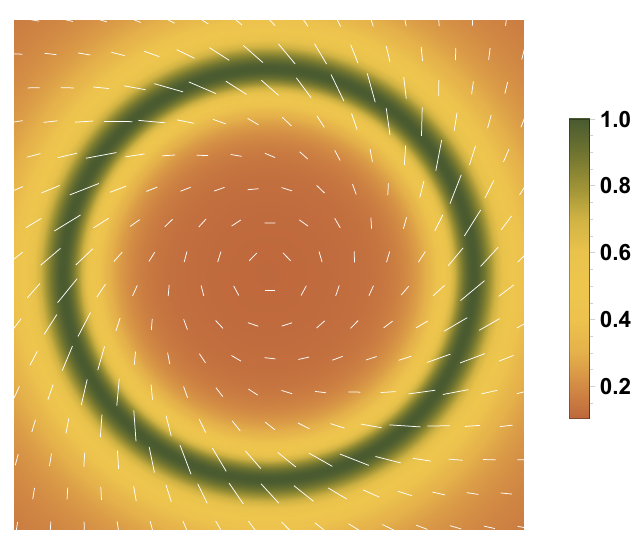}}
	\hfill
	\subfigure[\scriptsize $\xi=-0.05,\theta=30^\circ$]{%
		\includegraphics[width=0.24\textwidth]{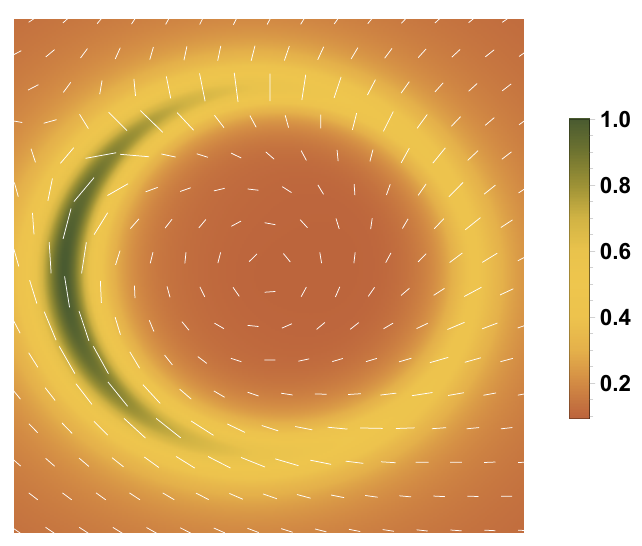}}
	\hfill
	\subfigure[\scriptsize $\xi=-0.05,\theta=60^\circ$]{%
		\includegraphics[width=0.24\textwidth]{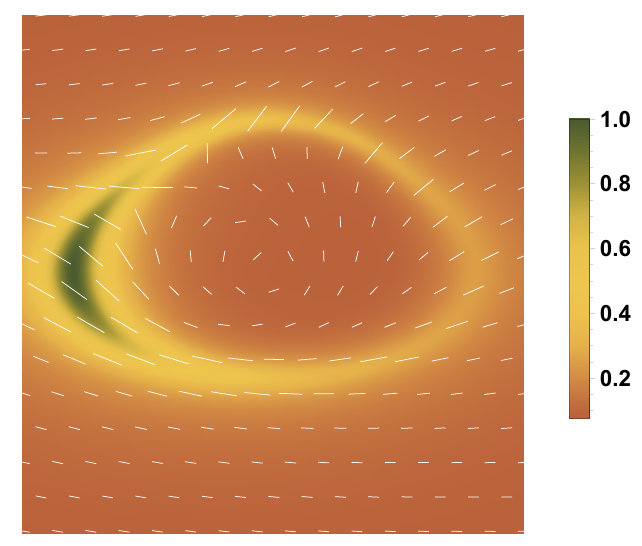}}
	\hfill
	\subfigure[\scriptsize $\xi=-0.05,\theta=86^\circ$]{%
		\includegraphics[width=0.24\textwidth]{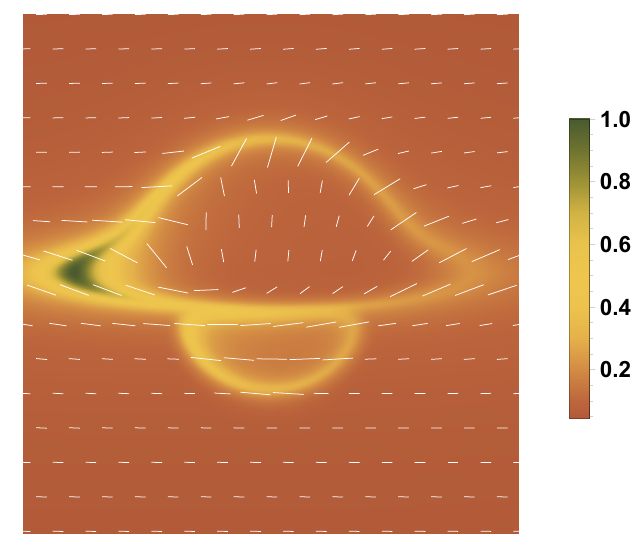}}
	
	\vspace{2mm}
	
	% Row 3
	\subfigure[\scriptsize $\xi=0.01,\theta=0^\circ$]{%
		\includegraphics[width=0.24\textwidth]{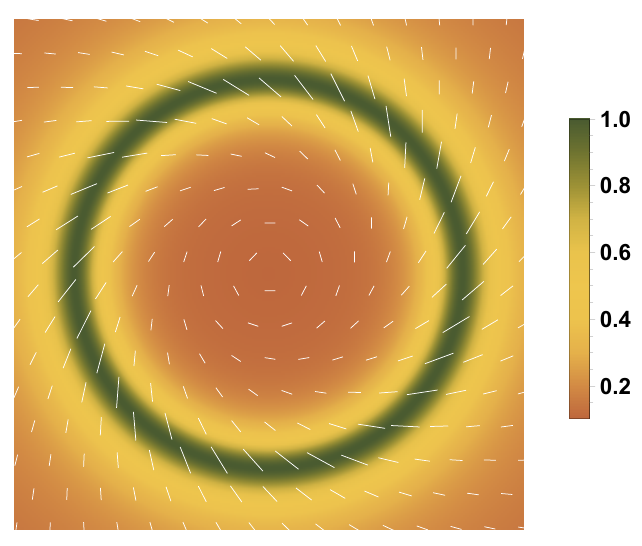}}
	\hfill
	\subfigure[\scriptsize $\xi=0.01,\theta=30^\circ$]{%
		\includegraphics[width=0.24\textwidth]{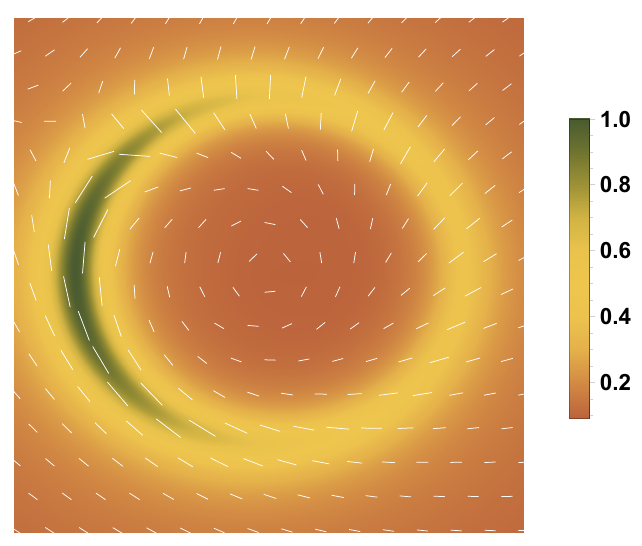}}
	\hfill
	\subfigure[\scriptsize $\xi=0.01,\theta=60^\circ$]{%
		\includegraphics[width=0.24\textwidth]{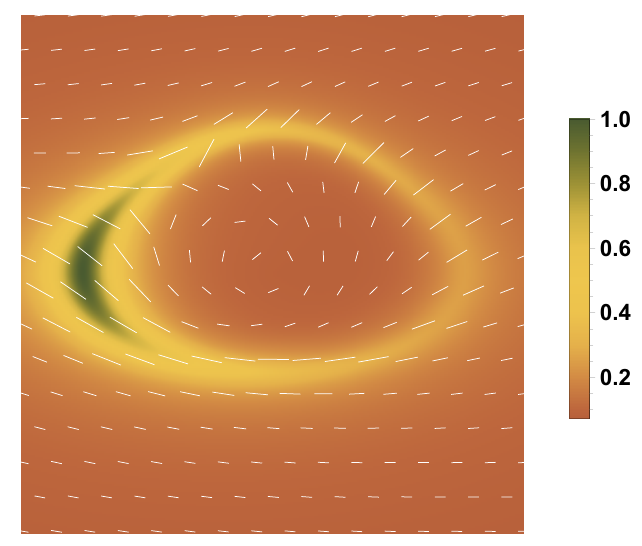}}
	\hfill
	\subfigure[\scriptsize $\xi=0.01,\theta=86^\circ$]{%
		\includegraphics[width=0.24\textwidth]{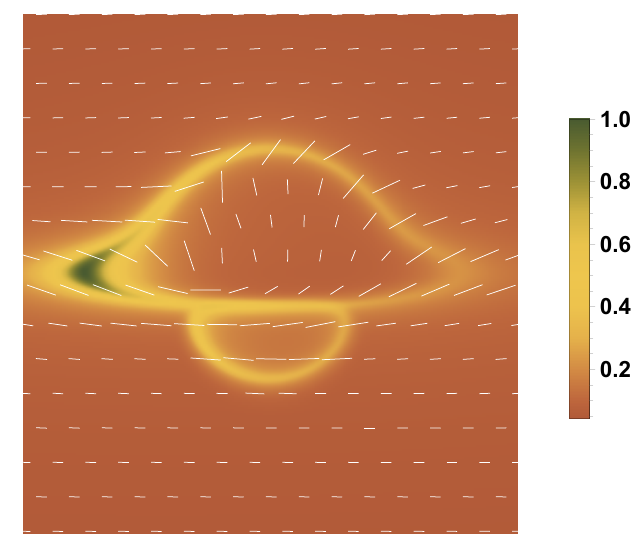}}
	
	\vspace{2mm}
	
	% Row 4
	\subfigure[\scriptsize $\xi=0.1,\theta=0^\circ$]{%
		\includegraphics[width=0.24\textwidth]{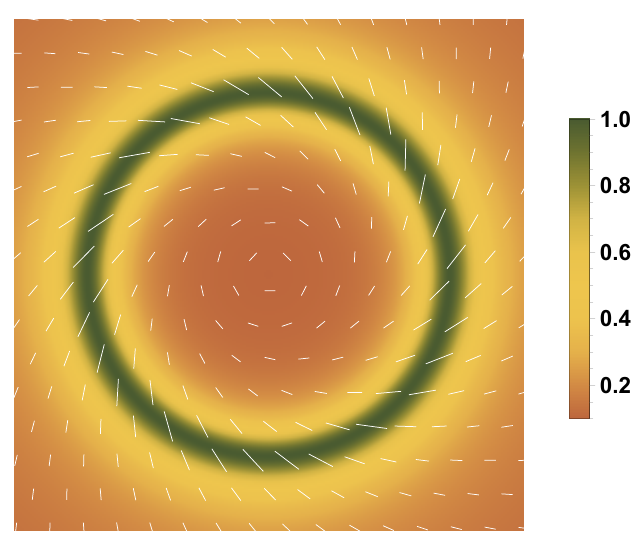}}
	\hfill
	\subfigure[\scriptsize $\xi=0.1,\theta=30^\circ$]{%
		\includegraphics[width=0.24\textwidth]{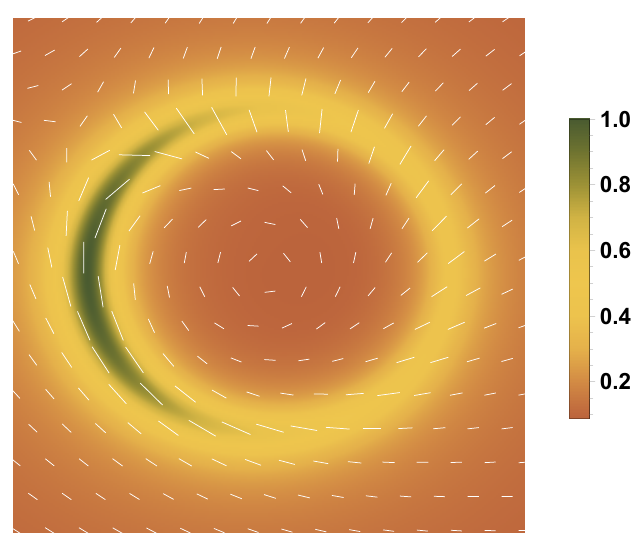}}
	\hfill
	\subfigure[\scriptsize $\xi=0.1,\theta=60^\circ$]{%
		\includegraphics[width=0.24\textwidth]{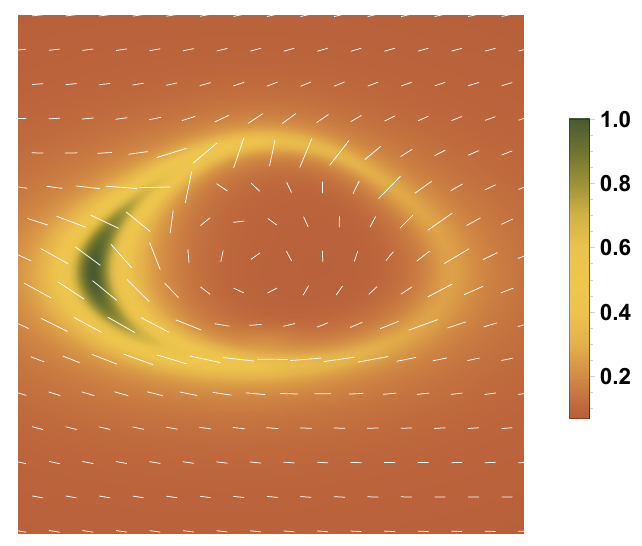}}
	\hfill
	\subfigure[\scriptsize $\xi=0.1,\theta=86^\circ$]{%
		\includegraphics[width=0.24\textwidth]{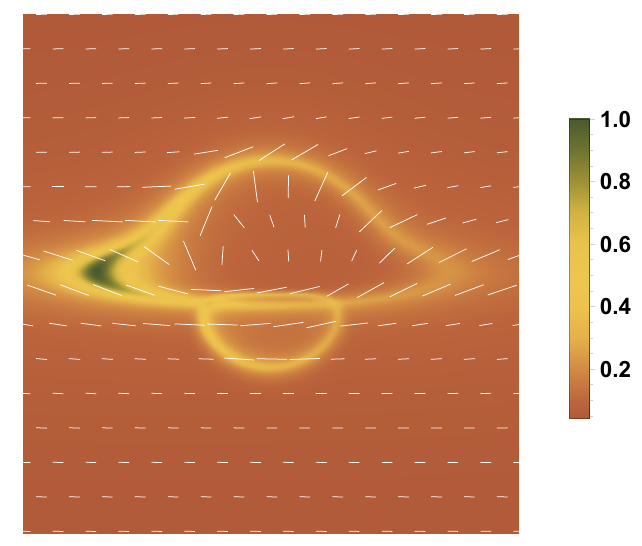}}
	
	\vspace{2mm}
	
	\caption{\label{fig14} The effect of the gravitational coupling parameter $\xi$ on the polarization image. Columns from left to right denote models with inclination angles $\theta = 0^\circ, 30^\circ, 60^\circ, 86^\circ$, and rows from top to bottom denote $\xi = -0.1, -0.05, 0.01, 0.1$.}
\end{figure*}

The optical images of various boson stars with a thin accretion disk are presented in Figure~\ref{fig13}. We set $\alpha_{\text{fov}} = 2.8^\circ$, $\psi_0 = 0.18$, with the observer located at $r_{\text{obs}} = 200$. It can be observed that for observation angles $\theta = 0^\circ, 30^\circ, 60^\circ$ (the first, second, and third columns), the optical image displays only the direct image, where $\xi$ affects the size but not the shape of the direct image. As $\xi$ increases, the direct image gradually becomes smaller. When $\theta = 86^\circ$ (the fourth column), a lensed image appears in the optical image of the boson star. The upper hat-shaped optical image corresponds to the direct image, while the lower D-shaped optical image represents the lensed image. At smaller observation angles, the direct image of the boson star forms a ring, primarily influenced by gravitational redshift. As the angle increases, the left side of the image becomes brighter than the right due to Doppler effects. The corresponding polarization image for Figure~\ref{fig13} is shown in Figure~\ref{fig14}. From the figure, it is evident that both $\xi$ and $\theta$ significantly affect the polarization vector. This indicates that combining optical images with polarization effects provides a more comprehensive way to reveal the radiation characteristics and spacetime structure around boson stars.

%%%%%%%%%%%%%%%%%%%%%%%%%%%%%%%%%%%%%%%%%%%%%%

\section{Conclusion and discussion}\label{sub8}
In this work, we have investigated the optical properties of mini boson stars within the framework of Palatini $f(R)$ gravity, adopting the quadratic form $f(R) = R + \xi R^2$. By numerically solving the modified field equations, we have analyzed photon trajectories and optical images generated under spherical light sources and thin accretion disks. The results reveal several key distinctions between boson stars and Schwarzschild black holes, providing theoretical insights into the astrophysical observability of exotic compact objects.

First, we find that boson stars lack photon rings, as the derivative of the effective potential does not vanish anywhere, in sharp contrast to black holes. This characteristic leads to optical images of boson stars being dominated by direct emissions from photons completing a single orbit. The absence of photon rings results in distinctive imaging features that could serve as potential observational signatures.

Second, we have demonstrated that the optical characteristics of boson stars are highly sensitive to the initial scalar field $\psi_0$ and the gravitational coupling parameter $\xi$. Variations in $\psi_0$ significantly affect the size of the boson star and the corresponding Einstein ring, while changes in $\xi$ influence the effective potential and redshift distributions. For polarization images, the total linear polarization intensity $\mathsf{P}_o$ is strongly dependent on the intensity. Unlike black holes, polarization vectors also appear inside the boson star. These dependencies highlight the critical role of model parameters in determining the observational properties of boson stars.

Our findings underscore the importance of high-resolution observations for distinguishing boson stars from black holes. The unique optical features of boson stars, such as the absence of photon rings and the distribution of polarization vectors, provide a promising pathway for probing the existence of exotic compact objects. Moreover, the use of Palatini $f(R)$ gravity in this context extends the applicability of modified gravity theories to compact astrophysical systems, offering an alternative to general relativity for exploring deviations in strong-field regimes.

Future research directions include extending the current analysis to include rotating boson stars and exploring the impact of higher-order terms in the $f(R)$ function. Additionally, incorporating more realistic accretion disk models and analyzing multi-wavelength observations could further improve the ability to distinguish boson stars from black holes. These efforts will enhance our understanding of exotic compact objects and contribute to the broader quest for observational tests of alternative gravity theories.

\section*{Data Availability Statement}
The data that support the findings of this study are available from the corresponding author upon reasonable request.

%%%%%%%%%%%%%%%%%%%%%%%%%%%%%%%%%%%%%%%%%%%%%%%%%%%%%%%%%%%%%%%%%%%%%%%%
%%%%%%%%%%%%%%%%%%%%%%%%%%%%%%%%%%%%%%%%%%%%%%%%%%%%%%%%%%%%%%%%%%%%%%%%
%%%%%%%%%%%%%%%%%%%%%%%%%%%%%%%%%%%%%%%%%%%%%%%%%%%%%%%%%%%%%%%%%%%%%%%%

\cleardoublepage

\vspace{10pt}
\noindent {\bf Acknowledgments}

\noindent
This work is supported by the National Natural Science Foundation of China (Grant Nos. 11875095 and 11903025), by the Natural Science Foundation of Chongqing (CSTB2023NSCQ MSX0594), and the Fund Project of Chongqing Normal University (Grant Number: 24XLB033).

% BibTeX
\bibliographystyle{JHEP} % 指定参考文献格式
\bibliography{biblio} % 引用 BibTeX 文件

@article{afonso2019correspondence,
	title={Correspondence between modified gravity and general relativity with scalar fields},
	author={Afonso, Victor I and Olmo, Gonzalo J and Orazi, Emanuele and Rubiera-Garcia, Diego},
	journal={Physical Review D},
	volume={99},
	number={4},
	pages={044040},
	year={2019},
	publisher={APS}
}

@article{liebling2023dynamical,
	title={Dynamical boson stars},
	author={Liebling, Steven L and Palenzuela, Carlos},
	journal={Living Reviews in Relativity},
	volume={26},
	number={1},
	pages={1},
	year={2023},
	publisher={Springer}
}

@article{herdeiro2017asymptotically,
	title={Asymptotically flat scalar, Dirac and Proca stars: discrete vs. continuous families of solutions},
	author={Herdeiro, Carlos AR and Pombo, Alexandre M and Radu, Eugen},
	journal={Physics Letters B},
	volume={773},
	pages={654--662},
	year={2017},
	publisher={Elsevier}
}

@article{gralla2020shape,
	title={The shape of the black hole photon ring: A precise test of strong-field general relativity},
	author={Gralla, Samuel E and Lupsasca, Alexandru and Marrone, Daniel P},
	journal={Physical Review D},
	volume={102},
	number={12},
	pages={124004},
	year={2020},
	publisher={APS}
}

@article{vincent2022images,
	title={Images and photon ring signatures of thick disks around black holes},
	author={Vincent, Frederic H and Gralla, Samuel E and Lupsasca, Alexandru and Wielgus, Maciek},
	journal={Astronomy \& Astrophysics},
	volume={667},
	pages={A170},
	year={2022},
	publisher={EDP Sciences}
}

@article{ligo2019gwtc,
	title={GWTC-1: A Gravitational-Wave Transient Catalog of Compact Binary Mergers Observed by LIGO and Virgo during the First and Second Observing Runs},
	author={LIGO Scientific Collaboration and Virgo Collaboration and others},
	year={2019}
}

@article{abbott2021gwtc,
	title={GWTC-2: compact binary coalescences observed by LIGO and Virgo during the first half of the third observing run},
	author={Abbott, Richard and Abbott, TD and Abraham, S and Acernese, F and Ackley, K and Adams, A and Adams, C and Adhikari, RX and Adya, VB and Affeldt, Christoph and others},
	journal={Physical Review X},
	volume={11},
	number={2},
	pages={021053},
	year={2021},
	publisher={APS}
}

@article{abbott2020gw190814,
	title={GW190814: Gravitational waves from the coalescence of a 23 solar mass black hole with a 2.6 solar mass compact object},
	author={Abbott, Richard and Abbott, TD and Abraham, S and Acernese, Fausto and Ackley, K and Adams, C and Adhikari, Rana X and Adya, VB and Affeldt, Christoph and Agathos, Michail and others},
	journal={The Astrophysical Journal Letters},
	volume={896},
	number={2},
	pages={L44},
	year={2020},
	publisher={IOP Publishing}
}

@article{abbott2020gw190521,
	title={GW190521: a binary black hole merger with a total mass of 150 $M_\odot$},
	author={Abbott, Richard and Abbott, TD and Abraham, S and Acernese, Fausto and Ackley, K and Adams, C and Adhikari, RX and Adya, VB and Affeldt, Christoph and Agathos, M and others},
	journal={Physical review letters},
	volume={125},
	number={10},
	pages={101102},
	year={2020},
	publisher={APS}
}

@article{abbott2020properties,
	title={Properties and astrophysical implications of the 150 $M_\odot$ binary black hole merger GW190521},
	author={Abbott, Richard and Abbott, TD and Abraham, S and Acernese, F and Ackley, K and Adams, C and Adhikari, RX and Adya, VB and Affeldt, C and Agathos, M and others},
	journal={The Astrophysical Journal Letters},
	volume={900},
	number={1},
	pages={L13},
	year={2020},
	publisher={IOP Publishing}
}

@article{de2021gw190521,
	title={GW190521 mass gap event and the primordial black hole scenario},
	author={De Luca, V and Desjacques, V and Franciolini, G and Pani, P and Riotto, A},
	journal={Physical review letters},
	volume={126},
	number={5},
	pages={051101},
	year={2021},
	publisher={APS}
}

@article{bustillo2021gw190521,
	title={GW190521 as a merger of Proca stars: a potential new vector boson of 8.7$\times$ 10-13 eV},
	author={Bustillo, Juan Calder{\'o}n and Sanchis-Gual, Nicolas and Torres-Forn{\'e}, Alejandro and Font, Jos{\'e} A and Vajpeyi, Avi and Smith, Rory and Herdeiro, Carlos and Radu, Eugen and Leong, Samson HW},
	journal={Physical Review Letters},
	volume={126},
	number={8},
	pages={081101},
	year={2021},
	publisher={APS}
}

@article{cardoso2019testing,
	title={Testing the nature of dark compact objects: a status report},
	author={Cardoso, Vitor and Pani, Paolo},
	journal={Living Reviews in Relativity},
	volume={22},
	number={1},
	pages={4},
	year={2019},
	publisher={Springer}
}

@article{herdeiro2021imitation,
	title={The imitation game: Proca stars that can mimic the Schwarzschild shadow},
	author={Herdeiro, Carlos AR and Pombo, Alexandre M and Radu, Eugen and Cunha, Pedro VP and Sanchis-Gual, Nicolas},
	journal={Journal of Cosmology and Astroparticle Physics},
	volume={2021},
	number={04},
	pages={051},
	year={2021},
	publisher={IOP Publishing}
}

@article{kaup1968klein,
	title={Klein-gordon geon},
	author={Kaup, David J},
	journal={Physical Review},
	volume={172},
	number={5},
	pages={1331},
	year={1968},
	publisher={APS}
}

@article{ruffini1969systems,
	title={Systems of self-gravitating particles in general relativity and the concept of an equation of state},
	author={Ruffini, Remo and Bonazzola, Silvano},
	journal={Physical Review},
	volume={187},
	number={5},
	pages={1767},
	year={1969},
	publisher={APS}
}

@article{di2020dynamical,
	title={Dynamical bar-mode instability in spinning bosonic stars},
	author={Di Giovanni, Fabrizio and Sanchis-Gual, Nicolas and Cerd{\'a}-Dur{\'a}n, Pablo and Zilh{\~a}o, Miguel and Herdeiro, Carlos and Font, Jos{\'e} A and Radu, Eugen},
	journal={Physical Review D},
	volume={102},
	number={12},
	pages={124009},
	year={2020},
	publisher={APS}
}

@article{palenzuela2007head,
	title={Head-on collisions of boson stars},
	author={Palenzuela, Carlos and Olabarrieta, I and Lehner, L and Liebling, Steven L},
	journal={Physical Review D—Particles, Fields, Gravitation, and Cosmology},
	volume={75},
	number={6},
	pages={064005},
	year={2007},
	publisher={APS}
}

@article{palenzuela2008orbital,
	title={Orbital dynamics of binary boson star systems},
	author={Palenzuela, Carlos and Lehner, L and Liebling, Steven L},
	journal={Physical Review D—Particles, Fields, Gravitation, and Cosmology},
	volume={77},
	number={4},
	pages={044036},
	year={2008},
	publisher={APS}
}

@article{schunck1998rotating,
	title={Rotating boson star as an effective mass torus in general relativity},
	author={Schunck, Franz E and Mielke, Eckehard W},
	journal={Physics Letters A},
	volume={249},
	number={5-6},
	pages={389--394},
	year={1998},
	publisher={Elsevier}
}

@article{yoshida1997rotating,
	title={Rotating boson stars in general relativity},
	author={Yoshida, Shijun and Eriguchi, Yoshiharu},
	journal={Physical Review D},
	volume={56},
	number={2},
	pages={762},
	year={1997},
	publisher={APS}
}

@article{kleihaus2008rotating,
	title={Rotating boson stars and Q-balls. II. Negative parity and ergoregions},
	author={Kleihaus, Burkhard and Kunz, Jutta and List, Meike and Schaffer, Isabell},
	journal={Physical Review D—Particles, Fields, Gravitation, and Cosmology},
	volume={77},
	number={6},
	pages={064025},
	year={2008},
	publisher={APS}
}

@article{bezares2017final,
	title={Final fate of compact boson star mergers},
	author={Bezares, Miguel and Palenzuela, Carlos and Bona, Carles},
	journal={Physical Review D},
	volume={95},
	number={12},
	pages={124005},
	year={2017},
	publisher={APS}
}

@article{palenzuela2017gravitational,
	title={Gravitational wave signatures of highly compact boson star binaries},
	author={Palenzuela, Carlos and Pani, Paolo and Bezares, Miguel and Cardoso, Vitor and Lehner, Luis and Liebling, Steven},
	journal={Physical Review D},
	volume={96},
	number={10},
	pages={104058},
	year={2017},
	publisher={APS}
}

@article{olivares2020tell,
	title={How to tell an accreting boson star from a black hole},
	author={Olivares, Hector and Younsi, Ziri and Fromm, Christian M and De Laurentis, Mariafelicia and Porth, Oliver and Mizuno, Yosuke and Falcke, Heino and Kramer, Michael and Rezzolla, Luciano},
	journal={Monthly Notices of the Royal Astronomical Society},
	volume={497},
	number={1},
	pages={521--535},
	year={2020},
	publisher={Oxford University Press}
}

@article{vincent2016imaging,
	title={Imaging a boson star at the Galactic center},
	author={Vincent, FH and Meliani, Z and Grandcl{\'e}ment, P and Gourgoulhon, E and Straub, O},
	journal={Classical and Quantum Gravity},
	volume={33},
	number={10},
	pages={105015},
	year={2016},
	publisher={IOP Publishing}
}

@article{rosa2022shadows,
	title={Shadows of boson and Proca stars with thin accretion disks},
	author={Rosa, Jo{\~a}o Lu{\'\i}s and Rubiera-Garcia, Diego},
	journal={Physical Review D},
	volume={106},
	number={8},
	pages={084004},
	year={2022},
	publisher={APS}
}

@article{rosa2023imaging,
	title={Imaging compact boson stars with hot spots and thin accretion disks},
	author={Rosa, Jo{\~a}o Lu{\'\i}s and Macedo, Caio FB and Rubiera-Garcia, Diego},
	journal={Physical Review D},
	volume={108},
	number={4},
	pages={044021},
	year={2023},
	publisher={APS}
}

@article{guzman2006accretion,
	title={Accretion disk onto boson stars: A way to supplant black hole candidates},
	author={Guzm{\'a}n, F Siddhartha},
	journal={Physical Review D—Particles, Fields, Gravitation, and Cosmology},
	volume={73},
	number={2},
	pages={021501},
	year={2006},
	publisher={APS}
}

@article{olmo2011palatini,
	title={Palatini approach to modified gravity: f (R) theories and beyond},
	author={Olmo, Gonzalo J},
	journal={International Journal of Modern Physics D},
	volume={20},
	number={04},
	pages={413--462},
	year={2011},
	publisher={World Scientific}
}

@book{harko2018extensions,
	title={Extensions of f (R) Gravity: Curvature-Matter Couplings and Hybrid Metric-Palatini Theory},
	author={Harko, Tiberiu and Lobo, Francisco SN},
	volume={1},
	year={2018},
	publisher={Cambridge University Press}
}

@article{hehl1995metric,
	title={Metric-affine gauge theory of gravity: field equations, Noether identities, world spinors, and breaking of dilation invariance},
	author={Hehl, Friedrich W and McCrea, J Dermott and Mielke, Eckehard W and Ne'eman, Yuval},
	journal={Physics Reports},
	volume={258},
	number={1-2},
	pages={1--171},
	year={1995},
	publisher={Elsevier}
}

@article{beltran2019geometrical,
	title={The geometrical trinity of gravity},
	author={Beltran Jimenez, Jose and Heisenberg, Lavinia and Koivisto, Tomi S},
	journal={Universe},
	volume={5},
	number={7},
	pages={173},
	year={2019},
	publisher={MDPI}
}

@article{delhom2019ricci,
	title={Ricci-Based Gravity theories and their impact on Maxwell and nonlinear electromagnetic models},
	author={Delhom, Adria and Olmo, Gonzalo J and Orazi, Emanuele},
	journal={Journal of High Energy Physics},
	volume={2019},
	number={11},
	pages={1--24},
	year={2019},
	publisher={Springer}
}

@article{afonso2018mapping,
	title={Mapping nonlinear gravity into General Relativity with nonlinear electrodynamics},
	author={Afonso, Victor I and Olmo, Gonzalo J and Orazi, Emanuele and Rubiera-Garcia, Diego},
	journal={The European Physical Journal C},
	volume={78},
	pages={1--11},
	year={2018},
	publisher={Springer}
}

@article{afonso2018mapping2,
	title={Mapping Ricci-based theories of gravity into general relativity},
	author={Afonso, VI and Olmo, Gonzalo J and Rubiera-Garcia, D},
	journal={Physical Review D},
	volume={97},
	number={2},
	pages={021503},
	year={2018},
	publisher={APS}
}

@article{afonso2019new,
	title={New scalar compact objects in Ricci-based gravity theories},
	author={Afonso, Victor I and Olmo, Gonzalo J and Orazi, Emanuele and Rubiera-Garcia, Diego},
	journal={Journal of Cosmology and Astroparticle Physics},
	volume={2019},
	number={12},
	pages={044},
	year={2019},
	publisher={IOP Publishing}
}

@article{maso2021boson,
	title={Boson stars in Palatini $ f (\mathcal{R}) $ gravity},
	author={Maso-Ferrando, Andreu and Sanchis-Gual, Nicolas and Font, Jose A and Olmo, Gonzalo J},
	journal={arXiv preprint arXiv:2103.15705},
	year={2021}
}

@article{zeng2025holographic,
	title={Holographic images of a charged black hole in Lorentz symmetry breaking massive gravity},
	author={Zeng, Xiao-Xiong and Li, Li-Fang and Li, Pan and Liang, Bo and Xu, Peng},
	journal={Science China Physics, Mechanics \& Astronomy},
	volume={68},
	number={2},
	pages={220412},
	year={2025},
	publisher={Springer}
}

@article{guo2024image,
	title={Image of the Kerr--Newman Black Hole Surrounded by a Thin Accretion Disk},
	author={Guo, Sen and Huang, Yu-Xiang and Liang, En-Wei and Liang, Yu and Jiang, Qing-Quan and Lin, Kai},
	journal={The Astrophysical Journal},
	volume={975},
	number={2},
	pages={237},
	year={2024},
	publisher={IOP Publishing}
}

@article{cui2024optical,
	title={Optical appearance of numerical black hole solutions in higher derivative gravity},
	author={Cui, Yu-Hao and Guo, Sen and Huang, Yu-Xiang and Liang, Yu and Lin, Kai},
	journal={The European Physical Journal C},
	volume={84},
	number={8},
	pages={772},
	year={2024},
	publisher={Springer}
}

@article{guo2024influence,
	title={Influence of quantum correction on the Schwarzschild black hole polarized image},
	author={Guo, Sen and Huang, Yu-Xiang and Liu, Kuan and Liang, En-Wei and Lin, Kai},
	journal={The European Physical Journal C},
	volume={84},
	number={6},
	pages={601},
	year={2024},
	publisher={Springer}
}

@article{he2024observational,
	title={Observational appearance and extra photon rings of an asymmetric thin-shell wormhole with a Bardeen profile},
	author={He, Ke-Jian and Luo, Zhi and Guo, Sen and Li, Guo-Ping},
	journal={Chinese Physics C},
	volume={48},
	number={6},
	pages={065105},
	year={2024},
	publisher={IOP Publishing}
}

@article{hou2024unique,
	title={Unique Imprint of Black Hole Spin on the Polarization of Near-Horizon Images},
	author={Hou, Yehui and Huang, Jiewei and Mizuno, Yosuke and Guo, Minyong and Chen, Bin},
	journal={arXiv preprint arXiv:2409.07248},
	year={2024}
}

@article{huang2024images,
	title={Images and flares of geodesic hot spots around a Kerr black hole},
	author={Huang, Jiewei and Zhang, Zhenyu and Guo, Minyong and Chen, Bin},
	journal={Physical Review D},
	volume={109},
	number={12},
	pages={124062},
	year={2024},
	publisher={APS}
}

@article{zhang2024imaging,
	title={Imaging thick accretion disks and jets surrounding black holes},
	author={Zhang, Zhenyu and Hou, Yehui and Guo, Minyong and Chen, Bin},
	journal={Journal of Cosmology and Astroparticle Physics},
	volume={2024},
	number={05},
	pages={032},
	year={2024},
	publisher={IOP Publishing}
}

@article{zeng2020shadows,
	title={Shadows and photon spheres with spherical accretions in the four-dimensional Gauss--Bonnet black hole},
	author={Zeng, Xiao-Xiong and Zhang, Hai-Qing and Zhang, Hongbao},
	journal={The European Physical Journal C},
	volume={80},
	pages={1--11},
	year={2020},
	publisher={Springer}
}

@article{yang2024shadow,
	title={Shadow Images of Ghosh-Kumar Rotating Black Hole Illuminated By Spherical Light Sources and Thin Accretion Disks},
	author={Yang, Chen-Yu and Aslam, M Israr and Zeng, Xiao-Xiong and Saleem, Rabia},
	journal={arXiv preprint arXiv:2411.11807},
	year={2024}
}

@article{zeng2022shadows,
	title={The shadows and observational appearance of a noncommutative black hole surrounded by various profiles of accretions},
	author={Zeng, Xiao-Xiong and Li, Guo-Ping and He, Ke-Jian},
	journal={Nuclear Physics B},
	volume={974},
	pages={115639},
	year={2022},
	publisher={Elsevier}
}

@article{li2025observational,
	title={Observational features of massive boson stars with thin disk accretion},
	author={Li, Guo-Ping and Wu, Meng-Qi and He, Ke-Jian and Jiang, Qing-Quan},
	journal={arXiv preprint arXiv:2505.14734},
	year={2025}
}

@article{huang2024coport,
	title={Coport: a new public code for polarized radiative transfer in a covariant framework},
	author={Huang, Jiewei and Zheng, Liheng and Guo, Minyong and Chen, Bin},
	journal={Journal of Cosmology and Astroparticle Physics},
	volume={2024},
	number={11},
	pages={054},
	year={2024},
	publisher={IOP Publishing}
}

@article{yang2025shadow,
  title={Shadow and Polarization Images of Rotating Black Holes in Kalb-Ramond Gravity Illuminated by Several Thick Accretion Disks},
  author={Yang, Chen-Yu and Ye, Huan and Zeng, Xiao-Xiong},
  journal={arXiv preprint arXiv:2510.21229},
  year={2025}
}

@article{wang2025imaging1,
  title={Imaging and polarization patterns of various thick disks around Kerr-MOG black holes},
  author={Wang, Xinyu and Ye, Huan and Zeng, Xiao-Xiong},
  journal={arXiv preprint arXiv:2511.09379},
  year={2025}
}

@article{wang2025imaging2,
  title={Imaging and Polarimetric Signatures of Konoplya-Zhidenko Black Holes with Various Thick Disk},
  author={Wang, Xinyu and Wang, Yukang and Zeng, Xiao-Xiong},
  journal={arXiv preprint arXiv:2510.17906},
  year={2025}
}

\end{document}